\documentclass[aip,graphicx]{revtex4-1}
\draft 

\usepackage{geometry}
\usepackage{array}
\usepackage{subcaption}
\usepackage{graphicx}
\usepackage{setspace}
 
\usepackage{upgreek}
\usepackage{amsmath}
\mathchardef\mhyphen="2D  

\begin{document}

This article may be downloaded for personal use only. Any other use requires prior permission of the author and AIP Publishing. This article appeared in Physics of Fluids, 33(1), p.011907. and may be found at https://doi.org/10.1063/5.0036095
\clearpage

\title[Accepted in Physics of Fluids 33, 011907 (2021); doi.org/10.1063/5.0036095]{Pressure distribution and flow dynamics in a nasal airway using a scale resolving simulation}

\author{James Van Strien}
\affiliation{Mechanical \& Automotive Engineering, School of Engineering, RMIT University, PO Box 71, Bundoora, 3083, Australia}
\author{Kendra Shrestha}
 \affiliation{Mechanical \& Automotive Engineering, School of Engineering, RMIT University, PO Box 71, Bundoora, 3083, Australia}
 \author{Sargon Gabriel}
 \affiliation{Manufacturing, Materials \& Mechatronics Engineering, School of Engineering, RMIT University, PO Box 71, Bundoora, 3083, Australia}
\author{Petros Lappas}
 \affiliation{Mechanical \& Automotive Engineering, School of Engineering, RMIT University, PO Box 71, Bundoora, 3083, Australia}
\author{David F Fletcher}
 \affiliation{School of Chemical and Biomolecular Engineering, The University of Sydney, NSW 2006 Australia}
 \author{Narinder Singh}
 \affiliation{Faculty of Health \& Medicine, The University of Sydney, NSW 2006, Australia}
 \altaffiliation{Department of Otolaryngology, Head and Neck Surgery, Westmead Hospital, Hawkesbury Rd, Westmead NSW 2145, Australia}

\author{Kiao Inthavong}%
 \email{kiao.inthavong@rmit.edu.au}
\affiliation{Mechanical \& Automotive Engineering, School of Engineering, RMIT University, Bundoora, Australia}%


\begin{abstract}
Airflow through the nasal cavity exhibits a wide variety of fluid dynamics
behaviour due to the intricacy of the nasal geometry.  The flow is naturally unsteady and perhaps turbulent, despite CFD in the literature that assumes a steady laminar flow. Time-dependent simulations can be used to generate detailed data with the potential to uncover new flow behaviour, although they are more computationally intensive compared with steady-state simulations. Furthermore, verification of CFD results has relied on reported pressure drop (e.g. nasal resistance) across the nasal airway although the geometries used are different. This study investigated the unsteady nature of inhalation at flow rates of 10, 15, 20, and 30 L/min. A scale resolving CFD simulation using a hybrid RANS-LES model was used and compared with experimental measurements of the pressure distribution and the overall pressure drop in the nasal cavity. The experimental results indicated a large pressure drop across the nasal valve, as well as across the nasopharynx with the latter attributed to a narrow cross-sectional area. At a flowrate of 30 L/min, the CFD simulations showed that the anterior half of the nasal cavity displayed dominantly laminar but disturbed flow behaviour in the form of velocity fluctuations.  The posterior half of the nasal cavity displayed turbulent activity, characterised by erratic fluctuating velocities, which was enhanced by the wider cross-sectional areas in the coronal plane. At 15L/min, the flow field was laminar dominant with very little disturbance confirming a steady-state laminar flow assumption is viable at this flow rate.
\end{abstract}

\maketitle 

\section{Introduction}
The nasal cavity functions to humidify and warm the air, and filter it from particulates, bacteria and pollen, all the while allowing safe odorant gases to reach the olfactory receptors to facilitate the sense of smell. However, it is these very intricacies of the nasal cavity that make it such a complicated organ to analyse. Although extensive computational \cite{Subramaniam1998, Elad1993, Keyhani1995} and experimental \citep{Hahn1993, Kelly2004, Cheng2001} studies into the airflow behaviour in the nasal cavity, some critical questions about nasal airflow remain unanswered, such as how particular variations in nasal airway geometry combined with breathing flow rate may alter flow patterns. Doorly et al.'s \cite{Doorly2008,Doorly2008a} works provided a review and a comprehensive discussion of the relationship between nasal geometry and flow behaviour of a nasal cavity, while also presenting their own experimental and CFD simulation results for three models. Qualitative images were used for insightful visualisation and discussions.

Computational Fluid Dynamics (CFD) is a useful platform to study airﬂow in the nasal cavity. Recently, it has been used as a non-invasive method to investigate flow behaviour \cite{Inthavong2020, Xu2020, Ormiskangas2020, DeGabory2020, Pourmehran2020,Moreddu2020, Borojeni2020, FrankIto2019, Radulesco2019, Huang2019}. \citet{Li2017} compared results using various turbulence models against the experimental work of \citet{Hahn1993} replicating the unilateral cavity geometry with the nasopharynx omitted. The results demonstrated that a laminar flow model was most suitable for 180 mL/s flow rate through a single nasal chamber (21.6 L/min for both chambers), and the LES model and DNS were better suited for unilateral flow rates of 560 mL/s and greater. This is consistent with many studies reported in the literature where a laminar flow assumption was used for flow rates of 15 L/min, or less \cite{Zhao2004, Subramaniam1998, Shang2015a, Garcia2007, Ge2012, Goodarzi-Ardakani2016}. For flow rates of $>20$ L/min, a turbulent flow field was assumed. Early studies applied RANS (Reynolds Averaged Navier-Stokes) models including $k\mhyphen \varepsilon$ \cite{Lindemann2005}, $k\mhyphen \omega$ \cite{Horschler2006, Wen2008} and $v^2\mhyphen f$ \cite{Inthavong2011}. \citet{Kleinstreuer2003} demonstrated that to capture a laminar-transitional turbulent flow in a human airway, a low-Reynolds-number (LRN) $k\mhyphen \omega$ model was suitable. This model was later adopted by other researchers \cite{Zhu2013, Kim2013a, Ito2017}.

In nearly all the cited studies among others\cite{Kim2013, Wang2016, Wakayama2016, Kim2014, Dong2018} constant inhalation flow rates were modelled under steady-state conditions, which are quick to solve and therefore useful for determining consistent flow characteristics among extensive population sample studies. However, the results are time-invariant flow fields, but due to the intricate geometry of the nasal cavity, it is expected that flow fluctuations occur from either separated flow or due to turbulence. In fact, flow in the nasal cavity is most likely to experience both laminar and turbulent regimes \cite{Doorly2008,Doorly2008a,Inthavong2019}. Transient simulations made with Large Eddy Simulations (LES) can resolve both flow disturbance and turbulence. For example, \citet{Calmet2016} presented unsteady flow dynamics during rapid inhalation; \citet{Lee2010} simulated the entire respiratory (inhalation-exhalation) cycle; and \citet{Liu2007} and \citet{Ghahramani2017} investigated micron-sized particle deposition to overcome the deficiencies of turbulence dispersion with RANS (Reynolds Averaged Navier--Stokes) turbulence models. However, these studies did not include corresponding experimental data for verification.

The scale up in computational costs from a steady-state laminar flow to LES can be prohibitive because of the requirement of a small filtering mesh element size, and the small time steps required to satisfy the criterion of $CFL < 1$. In addition, there are strict requirements for achieving a satisfactory near-wall mesh defined by the viscous wall units ($x^+$, $y^+$, $z^+$). While wall-normal distance $y^+$ needs to be kept $<1$ the spanwise and streamwise directions become as equally important in respiratory airflows since the flow is much more three-dimensional than those found in standard test cases such as channel flows. However, LES studies in the respiratory airflows rarely address these requirements. 

An alternative is the hybrid RANS-LES turbulence model that seeks to exploit the best features of the RANS and LES approaches and relaxes the strict requirements of the near-wall mesh. This modelling technique was developed initially for external flow (e.g., aerospace, automotive \cite{Menter2018}) but has more recently been seen to offer significant advantages in internal flows as well \cite{Brown2020}. The essential idea is to exploit both models in their optimal domain of application. As such a $k\mhyphen \omega$  SST model is employed close to the wall as it can capture the boundary layer well and has good separation prediction capabilities on a suitably refined near-wall mesh in SST.  On the other hand, LES is best suited to modelling flow away from the walls and can capture the dynamic behaviour of wakes and separated flows. These hybrid or scale resolving models have undergone much development, which has culminated in the Stress Blended Eddy Simulation (SBES) where the blending between the eddy viscosity and subgrid-scale viscosity is explicit. Given its increasing use and demonstrated benefits, it was seen as timely to apply the model to this important biomedical flow.

CFD studies require verification/validation of the solution, and this has manifested in the form of overall pressure drop (reported in various studies)\citep{Wen2008, Weinhold2004, Kelly2004}. The overall pressure drop is taken between the surrounding ambient air, and a posterior location in the nasal cavity, usually the nasopharynx exit, or choanae. \cite{Inthavong2014} proposed a pressure drop relationship based on Bernoulli's equation and despite its inviscid flow assumption, the relationship compared well with a data-fitted relationship.

In this study, we aim to characterise the fluid flow through a nasal airway at a constant flow rate using CFD. To verify the solution, experimental data (from the same nasal model) for pressure values at the lateral wall and nasal floor locations were extracted and used to compare directly with the CFD results. The overall pressure drop was compared with reported data in the literature. For the CFD study, a scale resolving simulation with a hybrid RANS-LES was used to characterise the flow at a constant flow rate of 15~L/min, and 30 L/min, where the latter is approximately the peak inspiratory flow during a respiratory cycle.

\section{Method}
\subsection{Nasal cavity model}
A human nasal cavity model was reconstructed from CT scans of a 48-year-old Asian male through segmentation techniques. The scan resolution was $512 \times 512$px with slice thickness of 0.5mm The nasal model (labelled \emph{NC04} and used in previous studies)\citep{Shang2015, Dong2016} was evaluated by a qualified clinician (Ear Nose Throat specialist) as having a deviated septum and a narrow nasopharynx outlet. The segmentation procedure was performed using 3D slicer. The DICOM files were imported into 3D slicer, and we apply a Laplacian sharpening filter to enhance the image registration. Thresholding Hounsfield (HU) values of $-$1024 to 550 were used. Segmentation of individual anatomical regions of interest was performed and exported as an STL (Stereolithography) file. The STL regions were imported into Ansys SpaceClaim where surface repair was conducted through a ‘shrinkwrap’ technique. The segmentation produced a 3D volume computer model, and this was manually refined by reducing the effects of noise and smoothing unrealistic regions. The sinuses were omitted and truncated at their respective ostia. External facial features were included to ensure realistic inhalation at the nostrils \citep{Shang2015a}. To prevent large flow gradients forming at the nasopharynx outlet boundary, an artificial extension (50~mm in length) was added. A 3D model of 1mm wall thickness was printed using \emph{Visijet SL Clear} transparent SLA resin. Externally protruding ports (2~mm in diameter) were added to the external walls to allow pressure transducers to record pressure values at specific locations. There were three ports along the nasal cavity floor, four on lateral walls of both sides of the nasal cavity (fourteen in total), and two additional posterior ports at the nasopharynx (refer to Fig.~\ref{fig:portlocations}).

\subsection{Experimental pressure measurement station}
The experimental setup is depicted in Fig.~\ref{fig:pressuredifferential}, which shows (a) a simplified schematic and (b) a photograph of the test station and setup. A linear actuator controlled the airflow with a VNH5019 Motor Driver Carrier and an Arduino Uno R3 programmable logic controller. The in-built Pulse-Width Modulation (PWM) function in the Arduino sent a signal to the motor driver to alter the output voltage to the linear actuator. The flow was modulated using a pressure regulator and flow adjustment needle valve and monitored with a TSI4043 flow meter was used to monitor the constant flow rate. The breathing flow rates evaluated were 10, 15, 20, 30~L/min which are typical values used in computational studies, and it is within range of a full sinusoidal respiration (inhale and exhale) profile with a period of 4~s and a tidal volume of 500~mL. Sensirion SDP-1000 differential pressure transducers were used to measure the pressure drop between any two ports; totalling 16 measurements per flow rate. After a stable constant flow rate was established, data acquisition was logged at a rate of $10$~Hz for ten seconds and then averaged.

\subsection{CFD modelling}
CFD simulations were performed for constant inhalation flow rates for 10 and 15~L/min inhalation rates with a laminar flow model \citep{Kim2013,Croce2006,Zhao2006}, while for a 30 L/min inhalation, a Stress-Blended Eddy Simulation (SBES) turbulence model was used. The SBES model applies a Reynolds Averaged Navier--Stokes (RANS) model in the near wall region, whilst resolving the flow away from the wall with a Large Eddy Simulation (LES). Hybrid RANS-LES models, such as the SBES model, overcome the LES limitation associated with the need for very fine meshes in the wall boundary layer by integrating to the wall using a RANS model. The $ k\text{-}\omega $ SST (Shear Stress Transport) model was used with the LES Wall-Adapting Local Eddy-Viscosity (WALE) model for the scale resolving component. The incompressible flow equations describing the conservation for mass and momentum are expressed as:
\begin{align}
    \frac{\partial}{\partial x_i} \left( \bar{u}_i\right) &= 0 \\
    \frac{\partial}{\partial t}\left(\rho \bar{u}_i\right) + \frac{\partial}{\partial x_j}  \left(\rho \bar{u}_i \bar{u}_j \right) &=  -\frac{\partial \bar{p}}{\partial x_i} + \frac{\partial \sigma_{ij}}{\partial x_j} + \frac{\partial \tau_{ij}}{\partial x_j} 
\end{align}
where $ u_i $ is the flow velocity vector,  $ \rho $ is the fluid density and $ p $ is the pressure, and $\sigma_{ij}$  is the stress tensor due to molecular viscosity. The overbar $ \bar{\varphi} $ on the scalar quantity $ \varphi $ denotes a Reynolds-averaging operation in the RANS formulation and a spatio-temporal filtering operation in the LES formulation. The turbulence stress tensor $ \tau_{ij} \equiv \rho \left( \overline{u_i u_j} - \bar{u}_i \bar{u}_j \right) $ for the SBES formulation is defined to blend between the Reynolds stress tensor $ \tau_{ij}^\text{RANS} $ for the RANS formulation and the subgrid-scale stress tensor $ \tau_{ij}^\text{LES} $ for the LES formulation, according to the blending function
\begin{equation}
    \tau_{ij} = f_s \tau_{ij}^\text{RANS} + \left( 1 - f_s \right) \tau_{ij}^\text{LES}
\end{equation}
where $ 0 \le f_s \le 1 $ is the shielding function. 

The modelling was performed using Ansys Fluent (version 2019R3). The equations were solved using the coupled solver, which ensures very strong coupling between the pressure and velocity fields. The convective terms were resolved using second-order bounded schemes, and time-stepping was performed using the bounded second-order implicit scheme. Gradients were calculated using the least-squares cell-based method, and the pressure was discretised using a second-order scheme. Default under-relaxation factors were employed. A laminar model was used for flow rates of 10 and 15 L/min. For 30 L/min the hybrid RANS-LES model was used where the hemisphere dome (see Fig \ref{fig:geom}a) was a pressure inlet with a turbulence intensity of 5\% and viscosity ratio of 10, while the nasopharynx outlet was a mass flow boundary condition.

\subsection{Meshing and Time Step Requirements}
The LES component of the SBES model required a sufficiently fine mesh to resolve the large eddies. The model was meshed with a polyhedral mesh on its surface, filled with hexahedrons in the bulk flow region and this mesh was connected to the inflation vi polyhedra. Eight prism layers were placed on the walls where the normalised wall distance $y^+<1$ on all walls. The strict requirements of normalised wall distances $x^+$, and $z^+$ were not required for the SBES model. Three meshed models were evaluated, as summarised in Table~\ref{tab:mesh}, where the hexahedral cell length was $\Delta = 0.3$ mm, $0.2$ mm and $0.15$ mm, with eight prism layers set with first cell heights of 10\% of $\Delta$. 

\begin{table}[!h]
	\centering
	\begin{tabular}{p{2cm} >{\centering\arraybackslash}p{2cm} >{\centering\arraybackslash}p{2cm} >{\centering\arraybackslash}p{2cm} >{\centering\arraybackslash}p{2cm} }
		Mesh & $N_{cells} (\times 10^6)$ & $\Delta$~(mm) & $\quad N_{pl} \quad$ & $\quad h_{pl}$~(mm) \\
		\hline
		Mesh1 & 3.7 & 0.30 & 8 & 0.03\\
		Mesh2 & 8.1 & 0.20 & 8 & 0.02\\
		Mesh3 & 11.8 & 0.15 & 8 & 0.015\\
		\hline
	\end{tabular}
	\caption{Summary of different mesh resolution parameters with
		$N_{cells}$:~number of mesh cells/elements, $\Delta$:~grid size, $N_{pl}$:~number of prism layers and $h_{pl}$: height of first prism layer.}
	\label{tab:mesh}
\end{table}

To ensure sufficient spatial resolution was achieved, the mesh was evaluated based on the turbulence integral length-scale which includes most of the energy containing eddies and is estimated from a precursor $k\text{-}\omega$ SST model simulation using
\begin{equation}
l_o = \frac{k^{1.5}}{\varepsilon}=\frac{k^{0.5}}{C_\mu \omega},	\hspace{15mm} \mathrm{where} \; C_\mu = 0.09
\end{equation}

Based on Kolmogorov’s energy spectrum, 80--90\% of the turbulence kinetic energy can be resolved by the largest eddies if approximately five to twelve cells in each direction are created for an eddy with scale $l_o$. The mesh analysis through contour plots of the coronal slices indicated that 5-10 cells were found. 

The time-step size also needs to resolve the energy-containing eddies. An estimate of the time resolution of the model was calculated based on the time scale defined as
\[
\lambda_g/u' = (15\nu/\varepsilon)^{1/2} = \sqrt{15}\tau_\eta
\]
where the minimum value was $\tau_\eta = 5.0 \times 10^{-5}$~s. A smaller value of $\tau_\eta =2.0 \times 10^{-5}$~s was used to ensure the time resolution was sufficient to establish a scale resolved flow field. 

A precursor steady-state RANS simulation was performed to establish the flowfield. The simulation was then switched to transient and simulated for 0.15~s to allow the scale resolving components to establish. Monitoring points were located on five coronal planes, and data sampling began recording for velocity, pressure and resolved and unresolved turbulence kinetic energy components. The monitoring points sampled for a period of 0.20~s, and the simulation ended after 0.35~s. The Reynolds and Womersley numbers near the nasopharynx were calculated as 1690, and 2.44 respectively based on the following values of: $D_h=0.0149$~m; $U=1.66$~m/s; $\rho=1.225$~kg/m$^3$; $\upmu=1.79\times 10^{-5}$~Pa.s; $\omega = 2\pi f = 1.57~\text{s}^{-1}$ based on on a nominal breathing frequency  corresponding to approximately 15 breaths per minute.

Non-dimensional values of the grid spacing was evaluated with the preliminary results and were based on the integral length scale calculated from the cross-sections shown in Fig.~\ref{fig:meanVals} giving normalised grid spacing of $\Delta ^+ = 0.583$. The turbulence time scale is estimated by $\tau_\eta = \sqrt{\nu/\varepsilon}$ giving $t^+ = 0.0536$.

\section{Results}
\subsection{Pressure Distribution and Pressure Drop Comparison}
The pressure differential between the ambient air and port locations from the experimental and CFD results are given in Fig.~\ref{fig:press-distn}. The plots combine the pressure distribution in the left (black colour), the right (red colour), and the posterior (blue colour) nasal cavity. The pressure differentials were slightly higher along the lateral walls (labelled L1 to L4) compared with the floor locations (labelled F1 to F3); while the right cavity (with a larger flow rate moving through) exhibited greater pressure differential values than the left cavity. 

A large pressure drop occurred between the posterior positions P1 and P2 caused by the severe narrowing of the nasopharynx cross-sectional area, which is not typically representative of the population. The cross-sectional area at the posterior slice (see Fig.~\ref{fig:geom}b) was $A_c = 1.737$~cm$^2$ which reduces to $A_c = 0.575$~cm$^2$, at the nasopharynx outlet (horizontal plane). Pressure drop across the nasal valve is best represented by the difference between locations F1 and F2, and this also showed a large value. The CFD results for the pressure distribution showed a generally good match with the experimental data where the discrepancies depicted CFD results with a higher pressure drop than the experimental values, particularly in the left cavity for the flow rate of 30 L/min.

Fig.~\ref{fig:pressDrop} shows the overall pressure drop between the ambient and the posterior locations P1 and P2, obtained from the CFD and experimental results. The results from the two posterior locations were given to demonstrate the sensitivity of the results when the nasopharynx exhibits a sharp reduction in cross-sectional area. A comparison with experimental data of \citet{Segal2008, Kelly2004, Weinhold2004} and computational results of \citet{Subramaniam1998,Wen2008, Ge2012, Inthavong2014} is shown which showed that the pressure drop between the ambient and location P1 was most consistent with the reported data, and follows the trend-line from \citet{Inthavong2014}. Each nasal geometry is unique, and therefore the overall pressure drop represents an overall flow resistance in their respective nasal geometry and neglects local influences.  

\subsection{Mean Velocity Contours}
The mean and RMSE (Root Mean Square Error) velocity magnitude contours were taken at multiple slices in the three anatomical planes, (i.e. sagittal, transverse, and coronal) shown in Fig.~\ref{fig:meanVals}. The RMSE is defined as the root mean square of the difference between the instantaneous velocity and the average velocity, given as
\[
RMSE = \sqrt{\frac{1}{N} \sum_{i=1}^{N} \left( u_i-\overline{u} \right )^2}
\]
which represents the spread of the velocity magnitude. It can be used as an index of comparison to identify where the main regions of instantaneous velocity deviated from mean results.

Two sagittal planes, one through the left and the other through the right cavities, demonstrate the streamwise flow direction, which therefore represents the main flow path. The flow accelerates through the nasal valve region directed at an approximate $45^\circ$ direction before impinging on the middle turbinate. The transverse planes show the flow paths in the lateral directions, where at mid-height (y-midplane) the flow separates into two streams with one moving laterally, and the other moving along the septal wall (denoted by the arrows in Fig.~\ref{fig:meanVals}c). The coronal planes showed high velocities occurring in the anterior plane and the bulk flow remaining in the inferior half and then the middle regions of the main nasal passage. The superior locations where the olfactory region lies, experience low flow velocities. Furthermore, the main deviations are found in the right cavity where most of the flow moves through due to the larger cross-sectional area. Locally, the deviations occur in the middle region of the `main-1' and `main-3' coronal planes, and the posterior plane.

\subsection{Flow Characteristics in the Nasal Cavity}
The instantaneous velocity contours were taken at $t=0.20$~s (after the flow had been established) and are shown in Fig.~\ref{fig:flowfea}a. Fluctuations were found in the nasopharynx, along the nasal floor, and the nasal septum wall. Using these contours as a basis of the flow field, velocity vectors are shown in Fig.~\ref{fig:flowfea}b and labelled $a$ to $e$ with key flow features. This included (a) flow entering the nostril that is directed at an approximate angle of $40^\circ$ and (b) the air impinges on the anterior septum where the majority of the air moves through the inferior half of the main nasal passage. (c) A recirculating flow was found past the vestibule notch caused by flow separation. (d) The flow that is directed superiorly also impinges onto the head of the middle turbinate. (e) Posteriorly, multiple flow streams mix, creating a highly turbulent and disturbed flow. (f) On the very lateral and posterior regions of the inferior turbinates, the vectors exhibit reverse flow where the flow is returning towards the front of the nose, although this is at a very low velocity.

Flow streamlines have been depicted in past studies either as sketches \citep{Keyhani1995, Schreck1993, Subramaniam1998} or extracted CFD images \citep{Wen2008, Inthavong2019} and are shown in Fig.~\ref{fig:stream}, which indicates some common flow features despite the different nasal cavity models. These include a disturbed flow in the vestibule; two streams that separate from the nasal valve region; a recirculating region in the superior anterior nasal notch, three streams converging at the choana to form a highly mixed region in the entrance into the nasopharynx. The pathlines in the current CFD study were taken at a particular time instant ($t = 0.20$ s), and this was able to reveal temporal disturbances in the labelled (red arrows, also see video in the Supplementary material) stream paths.

\subsection{Fluctuation Analysis}
Transient monitoring points located at each coronal slice recorded velocity, and turbulence kinetic energy over a time period $t =0.16$~s to $0.35$~s and is shown in Fig.~\ref{fig:fluct}. The velocity magnitudes in the `Anterior' and `Main-1' were of the order of 8~m/s in the right nasal cavity and 6 to 7~m/s in the left nasal cavity. The fluctuations were negligible in the left cavity. The fluctuations in the right cavity exhibited small amplitudes (peak of 0.15 m/s) and showed an ordered repeating pattern. The spectral analysis of the velocity fluctuation showed a dominant frequency occurring at $f = 184$~Hz, which is a manifestation of the flow separation from the upstream vestibule. \citet{Doorly2008} found flow separation in a more superior location displayed a regular
frequency of approximately 460 Hz. These regular frequencies confirm a laminar disturbed flow behaviour. The planes `Main-2, Main-3', and `Posterior' exhibited much lower velocity magnitudes but higher amplitudes and less well ordered repeating patterns. In these planes, there was an increase in the dominant frequencies, and in the `Posterior' plane, there was no order at all. 

A power spectral density analysis was created by determining the turbulence kinetic energy, $k$. The resolved component of $k$, produced by the large eddies, was obtained from the velocity fluctuations monitored at each point given in Fig.~\ref{fig:fluct}, over the monitoring period of 0.4~s. The unresolved $k$ produced by the small eddies (that were smaller than the sub-grid scale) were explicitly modelled. This unresolved turbulence component was assumed isotropic and calculated  by, $k_u = (C_w \Delta S)^2/0.3$, where $\Delta$ is the grid length scale and $C_w = 0.325$ in the WALE model. A fast-fourier-transform was applied to the data from each monitored point to produce the power spectrum. The spectral analysis of the turbulence kinetic energy identified how much of the inertial sub-range was resolved by comparing the gradient $-5/3$, which is associated with the range of frequencies in which the energy cascade takes place. It is dominated by inertial energy transfer, where energy production is equal to the energy dissipation. This behaviour was most evident and strongly correlated in the monitoring points found at Main-2, Main-3', and `Posterior' suggesting a strong presence of turbulence in this posterior half of the nasal cavity.

A similar analysis was performed for a lower flow rate of 15~L/min since at this flow rate, a laminar flow assumption, using the '\textit{laminar}' viscous model (absence of turbulence) is commonly adopted \citep{Zhu2011, Shi2006, Zhao2006} but without justification. Fig.~\ref{fig:fluct15} demonstrates a relatively constant flow with an absence of fluctuating velocities except for some minor changes in the posterior region. This suggests that the locations monitored exhibited very orderly flow behaviour that is characteristic of a laminar flow.

\section{Discussion}
Nasal airway resistance is deduced from the pressure drop by the relationship 
$R = \Delta P / Q$, where Q is the flowrate with units of $mL/s$ for unilateral airflow resistance, and $R_T = \left ( R_L + R_R \right )/ R_L R_R$ (parallel resistance form) for bilateral airflow resistance, where the subscripts $L$ and $R$ represent the left and right cavities, respectively. Our results produced resistance values of 0.03, 0.04, 0.05, 0.09 Pa$\cdot$s/mL for flow rates of 10, 15, 20, 30~L/min. Comparatively, \citet{Bruning2020} found a median value of 0.05 Pa$\cdot$s/mL  for 12L/min; and \citet{Borojeni2020} found a range of 0.017 to 0.12 Pa$\cdot$s/mL for bilateral resistance at 15L/min. 

A primary cause for resistance is undoubtedly constrictions in the geometry. The relationship between flow resistance and a characteristic area have been proposed by \citet{Rojas2018} (nostril area, $R= f(A_{nostril})$) and \citet{Garcia2016} (minimum cross-sectional area $R= f(A_{min})$). \citet{Garcia2016} found that when the airway constriction, defined by the minimum cross-sectional area, is severe enough the resistance at the constricted overwhelms the resistance of all other regions of the nasal cavity. Their model used a pressure drop of 19~Pa, which yielded approximately 15~L/min (250~mL/s). This correlates with the nasal valve region, which has been identified as the minimum cross-sectional area attributing to more than 50\% of total airway resistance \citep{Yu2008}. Additionally, \citet{Hildebrandt2013} and  \citet{Bruning2020} found a maximal pressure gradient (decreasing) at the isthmus nasi in an average nasal geometry from 25 patient models. The present study confirmed a high resistance across the nasal valve region (from location F1 to F2). The pressure drop across the nasal valve as a proportion of the overall pressure drop to posterior location P1 showed (see appendix A for table of values) its contribution as 56\% to 74\% in the left cavity and 37\% to 47\% in the right cavity for the flowrates investigated. The nasal valve has been implicated as a source of high flow resistance due to its reduced cross-sectional area producing a converging-diverging type passageway. There was also a significant pressure drop at the nasopharynx region, and this was due to the abnormally narrow cross-sectional area suggesting that patients with this condition are expected to experience high shear stress loadings in the back of the nose. A shift in nasal patency perception may be felt in the posterior nose rather than in the nasal valve region. The results demonstrated that a reduction in cross-sectional area in other regions would cause a secondary local region of high resistance. 

The RMSE values showed significant deviations of instantaneous flow relative to its mean flow occurring at the nasal valve exit and in the middle turbinate regions. These deviations suggest strong disturbed flow features not captured in time-averaged flows. Their presence is expected to have an enhanced effect on the air conditioning capacity of the nose, and particle dispersion and deposition in these local regions. There is also significant mixing in the nasopharynx due to the two air streams from the left and right cavities merging in a converging passageway.

Transient monitoring of data at points throughout the nasal cavity allowed more detailed information of the flow field than in a steady flow. The left nasal cavity was narrower, and the flow was laminar and exhibited negligible fluctuations. The right cavity was the more patent airway and showed higher velocities and fluctuations. In the anterior half, the monitored points exhibited cyclic fluctuations with a dominant frequency of 184~Hz (see animations in Supplementary material). Spectral analysis of the turbulence kinetic energy showed a weak correlation with the $-$5/3 slope suggesting a small contribution of turbulence in inertial sub-range the anterior half, and that the fluctuations were dominated by a disturbed laminar flow from flow separation. We note that only one monitoring point in the middle of the coronal plane was taken and that different levels of turbulence could exist in other regions. In the posterior half of the nasal cavity where the airway cross-section increased, the fluctuations were much more significant, the number of frequency modes increased, and the spectral analysis identified the turbulence energy cascade (inertial sub-range) region, to confirm the latter half of the nasal cavity exhibited very strong turbulent flow behaviour. 

At the lower flow rate of 15~L/min, researchers have commonly adopted a laminar flow assumption without justification. The spectral analysis of the transient velocity values monitored at nine locations indicates that a laminar model is justified for a steady simulation at 15~L/min. It was also evident that the larger cross-sectional area found at the nasopharynx is a significant contributor towards the onset of flow disturbances in the bulk flow region.

The results presented here were for a single model which is a limitation of the work where larger samples of data through multiple models would strengthen the work. Nevertheless, one consistent trend we found that is common regardless of different patient models is the constrictive nasal valve region and its significant contribution to airway resistance. Furthermore, the narrow nasopharynx (and in fact any constrictive regions) will be a further contributor towards airway resistance.

\section{Conclusion}
Experimental measurements determined the pressure distribution across the nasal walls, and this was matched very well with the CFD simulations. The pressure drop across the nasal valve was 56\% to 74\% in the left (congested) cavity and 37\% to 47\% in the right (patent) cavity, when the nasopharynx bend was excluded. The nasopharynx in this study exhibited an abnormally significant cross-sectional area reduction, also leading to a large pressure drop.  The nasal valve has been implicated as the primary contributor for airway resistance, due to its rapid decrease in cross-sectional area, and this phenomenon was consistent for the nasopharynx bend.

A transient scale resolving simulation using the hybrid RANS-LES $k\text{-}\omega$ SBES turbulence model was used to explore the flow features through the nasal cavity in detail. The results demonstrated separated flows leading to a mix of disturbed laminar to low-level turbulence flow in the anterior half of the nasal cavity, and a highly turbulent flow in the posterior half. As the flow entered the nostril, it was directed superiorly at an approximate angle of $40^\circ$, where it impinged on the septum and moved through the inferior half of the main nasal passageway before hitting the head of the middle turbinate. There was a recirculating flow above the vestibule notch due to flow separation. At the posterior nasal cavity, upstream jets converged into one intense mixing conduit before entering the nasopharynx. Instantaneous flow fields demonstrated temporal disturbances with fluctuating flow in the nasopharynx, along the nasal floor, and the nasal septum. 
A novelty of this work was the direct comparison of experimental measurements with a computational study that used a hybrid RANS-LES turbulence model, where we identified geometric features leading to flow separation and regions of laminar flow (in the anterior half) and turbulent flow behaviour (in the posterior half) of the nasal cavity. Larger model samples would strengthen these results in the future.

\section*{Supplementary Material}
\subsubsection*{S1: Velocity Fluctuation Animation}
\begin{enumerate}
	\item Animation 1: Velocity magnitudes at a flow rate of 30L/min in the coronal planes over the monitoring period that was analysed in Fig \ref{fig:fluct}
	\item Animation 2:  Velocity magnitudes at a flow rate of 30L/min in the sagittal planes over the monitoring period that was analysed in Fig \ref{fig:fluct}
\end{enumerate}

\subsubsection*{S2: Pressure Distribution Data}
Tabulated data of the measured and CFD simulated pressure distribution across the nasal cavity.

\subsubsection*{S3: CAD geometry of the nasal cavity for 3D printing}
A print-ready stl file of the 3D nasal cavity model with pressure ports depicted in Fig \ref{fig:geom}.

\subsubsection*{S4: CAD geometry of the nasal cavity used for CFD analysis}
An STL file of the 3D nasal cavity model ready for CFD meshing.

\begin{acknowledgments}
The authors gratefully acknowledge the financial support provided by the Garnett Passe and Rodney Williams Foundation Conjoint Grant 2019-22.
\end{acknowledgments}

\section*{Data Availability Statement}
The data that supports the findings of this study are available within the article [and its supplementary material].

\small
{\setstretch{1.0}
\bibliography{pressure}

\begin{thebibliography}{59}%
\makeatletter
\providecommand \@ifxundefined [1]{%
 \@ifx{#1\undefined}
}%
\providecommand \@ifnum [1]{%
 \ifnum #1\expandafter \@firstoftwo
 \else \expandafter \@secondoftwo
 \fi
}%
\providecommand \@ifx [1]{%
 \ifx #1\expandafter \@firstoftwo
 \else \expandafter \@secondoftwo
 \fi
}%
\providecommand \natexlab [1]{#1}%
\providecommand \enquote  [1]{``#1''}%
\providecommand \bibnamefont  [1]{#1}%
\providecommand \bibfnamefont [1]{#1}%
\providecommand \citenamefont [1]{#1}%
\providecommand \href@noop [0]{\@secondoftwo}%
\providecommand \href [0]{\begingroup \@sanitize@url \@href}%
\providecommand \@href[1]{\@@startlink{#1}\@@href}%
\providecommand \@@href[1]{\endgroup#1\@@endlink}%
\providecommand \@sanitize@url [0]{\catcode `\\12\catcode `\$12\catcode
  `\&12\catcode `\#12\catcode `\^12\catcode `\_12\catcode `\%12\relax}%
\providecommand \@@startlink[1]{}%
\providecommand \@@endlink[0]{}%
\providecommand \url  [0]{\begingroup\@sanitize@url \@url }%
\providecommand \@url [1]{\endgroup\@href {#1}{\urlprefix }}%
\providecommand \urlprefix  [0]{URL }%
\providecommand \Eprint [0]{\href }%
\providecommand \doibase [0]{http://dx.doi.org/}%
\providecommand \selectlanguage [0]{\@gobble}%
\providecommand \bibinfo  [0]{\@secondoftwo}%
\providecommand \bibfield  [0]{\@secondoftwo}%
\providecommand \translation [1]{[#1]}%
\providecommand \BibitemOpen [0]{}%
\providecommand \bibitemStop [0]{}%
\providecommand \bibitemNoStop [0]{.\EOS\space}%
\providecommand \EOS [0]{\spacefactor3000\relax}%
\providecommand \BibitemShut  [1]{\csname bibitem#1\endcsname}%
\let\auto@bib@innerbib\@empty
\bibitem [{\citenamefont {Subramaniam}\ \emph {et~al.}(1998)\citenamefont
  {Subramaniam}, \citenamefont {Richardson}, \citenamefont {Morgan},
  \citenamefont {Kimbell},\ and\ \citenamefont {Guilmette}}]{Subramaniam1998}%
  \BibitemOpen
  \bibfield  {author} {\bibinfo {author} {\bibfnamefont {R.}~\bibnamefont
  {Subramaniam}}, \bibinfo {author} {\bibfnamefont {R.}~\bibnamefont
  {Richardson}}, \bibinfo {author} {\bibfnamefont {K.}~\bibnamefont {Morgan}},
  \bibinfo {author} {\bibfnamefont {J.}~\bibnamefont {Kimbell}}, \ and\
  \bibinfo {author} {\bibfnamefont {R.}~\bibnamefont {Guilmette}},\ }\bibfield
  {title} {\enquote {\bibinfo {title} {{Computational fluid dynamics
  simulations of inspiratory airflow in the human nose and nasopharynx}},}\
  }\href {\doibase 10.1080/089583798197772} {\bibfield  {journal} {\bibinfo
  {journal} {Inhalation Toxicology}\ } (\bibinfo {year} {1998}),\
  10.1080/089583798197772}\BibitemShut {NoStop}%
\bibitem [{\citenamefont {Elad}\ \emph {et~al.}(1993)\citenamefont {Elad},
  \citenamefont {Liebenthal}, \citenamefont {Wenig},\ and\ \citenamefont
  {Einav}}]{Elad1993}%
  \BibitemOpen
  \bibfield  {author} {\bibinfo {author} {\bibfnamefont {D.}~\bibnamefont
  {Elad}}, \bibinfo {author} {\bibfnamefont {R.}~\bibnamefont {Liebenthal}},
  \bibinfo {author} {\bibfnamefont {B.~L.}\ \bibnamefont {Wenig}}, \ and\
  \bibinfo {author} {\bibfnamefont {S.}~\bibnamefont {Einav}},\ }\bibfield
  {title} {\enquote {\bibinfo {title} {{Analysis of air flow patterns in the
  human nose}},}\ }\href {\doibase 10.1007/BF02441806} {\bibfield  {journal}
  {\bibinfo  {journal} {Medical {\&} Biological Engineering {\&} Computing}\ }
  (\bibinfo {year} {1993}),\ 10.1007/BF02441806}\BibitemShut {NoStop}%
\bibitem [{\citenamefont {Keyhani}, \citenamefont {Scherer},\ and\
  \citenamefont {Mozell}(1995)}]{Keyhani1995}%
  \BibitemOpen
  \bibfield  {author} {\bibinfo {author} {\bibfnamefont {K.}~\bibnamefont
  {Keyhani}}, \bibinfo {author} {\bibfnamefont {P.~W.}\ \bibnamefont
  {Scherer}}, \ and\ \bibinfo {author} {\bibfnamefont {M.~M.}\ \bibnamefont
  {Mozell}},\ }\bibfield  {title} {\enquote {\bibinfo {title} {{Numerical
  simulation of airflow in the human nasal cavity}},}\ }\href {\doibase
  10.1115/1.2794204} {\bibfield  {journal} {\bibinfo  {journal} {Journal of
  Biomechanical Engineering}\ } (\bibinfo {year} {1995}),\
  10.1115/1.2794204}\BibitemShut {NoStop}%
\bibitem [{\citenamefont {Hahn}, \citenamefont {Scherer},\ and\ \citenamefont
  {Mozell}(1993)}]{Hahn1993}%
  \BibitemOpen
  \bibfield  {author} {\bibinfo {author} {\bibfnamefont {I.}~\bibnamefont
  {Hahn}}, \bibinfo {author} {\bibfnamefont {P.~W.}\ \bibnamefont {Scherer}}, \
  and\ \bibinfo {author} {\bibfnamefont {M.~M.}\ \bibnamefont {Mozell}},\
  }\bibfield  {title} {\enquote {\bibinfo {title} {{Velocity profiles measured
  for airflow through a large-scale model of the human nasal cavity}},}\ }\href
  {\doibase 10.1152/jappl.1993.75.5.2273} {\bibfield  {journal} {\bibinfo
  {journal} {Journal of Applied Physiology}\ } (\bibinfo {year} {1993}),\
  10.1152/jappl.1993.75.5.2273}\BibitemShut {NoStop}%
\bibitem [{\citenamefont {Kelly}\ \emph {et~al.}(2004)\citenamefont {Kelly},
  \citenamefont {Asgharian}, \citenamefont {Kimbell},\ and\ \citenamefont
  {Wong}}]{Kelly2004}%
  \BibitemOpen
  \bibfield  {author} {\bibinfo {author} {\bibfnamefont {J.~T.}\ \bibnamefont
  {Kelly}}, \bibinfo {author} {\bibfnamefont {B.}~\bibnamefont {Asgharian}},
  \bibinfo {author} {\bibfnamefont {J.~S.}\ \bibnamefont {Kimbell}}, \ and\
  \bibinfo {author} {\bibfnamefont {B.~A.}\ \bibnamefont {Wong}},\ }\bibfield
  {title} {\enquote {\bibinfo {title} {{Particle deposition in human nasal
  airway replicas manufactured by different methods. Part I: Inertial regime
  particles}},}\ }\href {\doibase 10.1080/027868290883360} {\bibfield
  {journal} {\bibinfo  {journal} {Aerosol Science and Technology}\ } (\bibinfo
  {year} {2004}),\ 10.1080/027868290883360}\BibitemShut {NoStop}%
\bibitem [{\citenamefont {Cheng}\ \emph {et~al.}(2001)\citenamefont {Cheng},
  \citenamefont {Holmes}, \citenamefont {Gao}, \citenamefont {Guilmette},
  \citenamefont {Li}, \citenamefont {Surakitbanharn},\ and\ \citenamefont
  {Rowlings}}]{Cheng2001}%
  \BibitemOpen
  \bibfield  {author} {\bibinfo {author} {\bibfnamefont {Y.}~\bibnamefont
  {Cheng}}, \bibinfo {author} {\bibfnamefont {T.}~\bibnamefont {Holmes}},
  \bibinfo {author} {\bibfnamefont {J.}~\bibnamefont {Gao}}, \bibinfo {author}
  {\bibfnamefont {R.}~\bibnamefont {Guilmette}}, \bibinfo {author}
  {\bibfnamefont {S.}~\bibnamefont {Li}}, \bibinfo {author} {\bibfnamefont
  {Y.}~\bibnamefont {Surakitbanharn}}, \ and\ \bibinfo {author} {\bibfnamefont
  {C.}~\bibnamefont {Rowlings}},\ }\bibfield  {title} {\enquote {\bibinfo
  {title} {{Characterization of nasal spray pumps and deposition pattern in a
  replica of the human nasal airway}},}\ }\href {\doibase
  10.1089/08942680152484199} {\bibfield  {journal} {\bibinfo  {journal}
  {Journal of aerosol medicine : the official journal of the International
  Society for Aerosols in Medicine}\ }\textbf {\bibinfo {volume} {14}},\
  \bibinfo {pages} {267--280} (\bibinfo {year} {2001})}\BibitemShut {NoStop}%
\bibitem [{\citenamefont {Doorly}\ \emph {et~al.}(2008)\citenamefont {Doorly},
  \citenamefont {Taylor}, \citenamefont {Gambaruto}, \citenamefont {Schroter},\
  and\ \citenamefont {Tolley}}]{Doorly2008}%
  \BibitemOpen
  \bibfield  {author} {\bibinfo {author} {\bibfnamefont {D.~J.}\ \bibnamefont
  {Doorly}}, \bibinfo {author} {\bibfnamefont {D.~J.}\ \bibnamefont {Taylor}},
  \bibinfo {author} {\bibfnamefont {A.~M.}\ \bibnamefont {Gambaruto}}, \bibinfo
  {author} {\bibfnamefont {R.~C.}\ \bibnamefont {Schroter}}, \ and\ \bibinfo
  {author} {\bibfnamefont {N.}~\bibnamefont {Tolley}},\ }\bibfield  {title}
  {\enquote {\bibinfo {title} {{Nasal architecture: Form and flow}},}\ }\href
  {\doibase 10.1098/rsta.2008.0083} {\bibfield  {journal} {\bibinfo  {journal}
  {Philosophical Transactions of the Royal Society A: Mathematical, Physical
  and Engineering Sciences}\ } (\bibinfo {year} {2008}),\
  10.1098/rsta.2008.0083}\BibitemShut {NoStop}%
\bibitem [{\citenamefont {Doorly}, \citenamefont {Taylor},\ and\ \citenamefont
  {Schroter}(2008)}]{Doorly2008a}%
  \BibitemOpen
  \bibfield  {author} {\bibinfo {author} {\bibfnamefont {D.~J.}\ \bibnamefont
  {Doorly}}, \bibinfo {author} {\bibfnamefont {D.~J.}\ \bibnamefont {Taylor}},
  \ and\ \bibinfo {author} {\bibfnamefont {R.~C.}\ \bibnamefont {Schroter}},\
  }\bibfield  {title} {\enquote {\bibinfo {title} {{Mechanics of airflow in the
  human nasal airways}},}\ }\href {\doibase 10.1016/j.resp.2008.07.027}
  {\bibfield  {journal} {\bibinfo  {journal} {Respiratory Physiology and
  Neurobiology}\ }\textbf {\bibinfo {volume} {163}},\ \bibinfo {pages}
  {100--110} (\bibinfo {year} {2008})}\BibitemShut {NoStop}%
\bibitem [{\citenamefont {Inthavong}(2020)}]{Inthavong2020}%
  \BibitemOpen
  \bibfield  {author} {\bibinfo {author} {\bibfnamefont {K.}~\bibnamefont
  {Inthavong}},\ }\bibfield  {title} {\enquote {\bibinfo {title} {{From indoor
  exposure to inhaled particle deposition: A multiphase journey of inhaled
  particles}},}\ }\href {\doibase 10.1007/s42757-019-0046-6} {\bibfield
  {journal} {\bibinfo  {journal} {Experimental and Computational Multiphase
  Flow}\ } (\bibinfo {year} {2020}),\ 10.1007/s42757-019-0046-6}\BibitemShut
  {NoStop}%
\bibitem [{\citenamefont {Xu}\ \emph {et~al.}(2020)\citenamefont {Xu},
  \citenamefont {Wu}, \citenamefont {Weng},\ and\ \citenamefont {Fu}}]{Xu2020}%
  \BibitemOpen
  \bibfield  {author} {\bibinfo {author} {\bibfnamefont {X.}~\bibnamefont
  {Xu}}, \bibinfo {author} {\bibfnamefont {J.}~\bibnamefont {Wu}}, \bibinfo
  {author} {\bibfnamefont {W.}~\bibnamefont {Weng}}, \ and\ \bibinfo {author}
  {\bibfnamefont {M.}~\bibnamefont {Fu}},\ }\bibfield  {title} {\enquote
  {\bibinfo {title} {{Investigation of inhalation and exhalation flow pattern
  in a realistic human upper airway model by PIV experiments and CFD
  simulations}},}\ }\href {\doibase 10.1007/s10237-020-01299-3} {\bibfield
  {journal} {\bibinfo  {journal} {Biomechanics and Modeling in Mechanobiology}\
  } (\bibinfo {year} {2020}),\ 10.1007/s10237-020-01299-3}\BibitemShut
  {NoStop}%
\bibitem [{\citenamefont {Ormiskangas}\ \emph {et~al.}(2020)\citenamefont
  {Ormiskangas}, \citenamefont {Valtonen}, \citenamefont {Kivek{\"{a}}s},
  \citenamefont {Dean}, \citenamefont {Poe}, \citenamefont {J{\"{a}}rnstedt},
  \citenamefont {Lekkala}, \citenamefont {Harju}, \citenamefont {Saarenrinne},\
  and\ \citenamefont {Rautiainen}}]{Ormiskangas2020}%
  \BibitemOpen
  \bibfield  {author} {\bibinfo {author} {\bibfnamefont {J.}~\bibnamefont
  {Ormiskangas}}, \bibinfo {author} {\bibfnamefont {O.}~\bibnamefont
  {Valtonen}}, \bibinfo {author} {\bibfnamefont {I.}~\bibnamefont
  {Kivek{\"{a}}s}}, \bibinfo {author} {\bibfnamefont {M.}~\bibnamefont {Dean}},
  \bibinfo {author} {\bibfnamefont {D.}~\bibnamefont {Poe}}, \bibinfo {author}
  {\bibfnamefont {J.}~\bibnamefont {J{\"{a}}rnstedt}}, \bibinfo {author}
  {\bibfnamefont {J.}~\bibnamefont {Lekkala}}, \bibinfo {author} {\bibfnamefont
  {T.}~\bibnamefont {Harju}}, \bibinfo {author} {\bibfnamefont
  {P.}~\bibnamefont {Saarenrinne}}, \ and\ \bibinfo {author} {\bibfnamefont
  {M.}~\bibnamefont {Rautiainen}},\ }\bibfield  {title} {\enquote {\bibinfo
  {title} {{Assessment of PIV performance in validating CFD models from nasal
  cavity CBCT scans}},}\ }\href {\doibase 10.1016/j.resp.2020.103508}
  {\bibfield  {journal} {\bibinfo  {journal} {Respiratory Physiology and
  Neurobiology}\ } (\bibinfo {year} {2020}),\
  10.1016/j.resp.2020.103508}\BibitemShut {NoStop}%
\bibitem [{\citenamefont {de~Gabory}\ \emph {et~al.}(2020)\citenamefont
  {de~Gabory}, \citenamefont {K{\'{e}}rimian}, \citenamefont {Baux},
  \citenamefont {Boisson},\ and\ \citenamefont {Bordenave}}]{DeGabory2020}%
  \BibitemOpen
  \bibfield  {author} {\bibinfo {author} {\bibfnamefont {L.}~\bibnamefont
  {de~Gabory}}, \bibinfo {author} {\bibfnamefont {M.}~\bibnamefont
  {K{\'{e}}rimian}}, \bibinfo {author} {\bibfnamefont {Y.}~\bibnamefont
  {Baux}}, \bibinfo {author} {\bibfnamefont {N.}~\bibnamefont {Boisson}}, \
  and\ \bibinfo {author} {\bibfnamefont {L.}~\bibnamefont {Bordenave}},\
  }\bibfield  {title} {\enquote {\bibinfo {title} {{Computational fluid
  dynamics simulation to compare large volume irrigation and continuous
  spraying during nasal irrigation}},}\ }\href {\doibase 10.1002/alr.22458}
  {\bibfield  {journal} {\bibinfo  {journal} {International Forum of Allergy
  and Rhinology}\ } (\bibinfo {year} {2020}),\ 10.1002/alr.22458}\BibitemShut
  {NoStop}%
\bibitem [{\citenamefont {Pourmehran}\ \emph {et~al.}(2020)\citenamefont
  {Pourmehran}, \citenamefont {Arjomandi}, \citenamefont {Cazzolato},
  \citenamefont {Ghanadi},\ and\ \citenamefont {Tian}}]{Pourmehran2020}%
  \BibitemOpen
  \bibfield  {author} {\bibinfo {author} {\bibfnamefont {O.}~\bibnamefont
  {Pourmehran}}, \bibinfo {author} {\bibfnamefont {M.}~\bibnamefont
  {Arjomandi}}, \bibinfo {author} {\bibfnamefont {B.}~\bibnamefont
  {Cazzolato}}, \bibinfo {author} {\bibfnamefont {F.}~\bibnamefont {Ghanadi}},
  \ and\ \bibinfo {author} {\bibfnamefont {Z.}~\bibnamefont {Tian}},\
  }\bibfield  {title} {\enquote {\bibinfo {title} {{The impact of geometrical
  parameters on acoustically driven drug delivery to maxillary sinuses}},}\
  }\href {\doibase 10.1007/s10237-019-01230-5} {\bibfield  {journal} {\bibinfo
  {journal} {Biomechanics and Modeling in Mechanobiology}\ } (\bibinfo {year}
  {2020}),\ 10.1007/s10237-019-01230-5}\BibitemShut {NoStop}%
\bibitem [{\citenamefont {Moreddu}\ \emph {et~al.}(2020)\citenamefont
  {Moreddu}, \citenamefont {Meister}, \citenamefont {Dabadie}, \citenamefont
  {Triglia}, \citenamefont {M{\'{e}}dale},\ and\ \citenamefont
  {Nicollas}}]{Moreddu2020}%
  \BibitemOpen
  \bibfield  {author} {\bibinfo {author} {\bibfnamefont {E.}~\bibnamefont
  {Moreddu}}, \bibinfo {author} {\bibfnamefont {L.}~\bibnamefont {Meister}},
  \bibinfo {author} {\bibfnamefont {A.}~\bibnamefont {Dabadie}}, \bibinfo
  {author} {\bibfnamefont {J.~M.}\ \bibnamefont {Triglia}}, \bibinfo {author}
  {\bibfnamefont {M.}~\bibnamefont {M{\'{e}}dale}}, \ and\ \bibinfo {author}
  {\bibfnamefont {R.}~\bibnamefont {Nicollas}},\ }\bibfield  {title} {\enquote
  {\bibinfo {title} {{Numerical simulation of nasal airflows and thermal air
  modification in newborns}},}\ }\href {\doibase 10.1007/s11517-019-02092-w}
  {\bibfield  {journal} {\bibinfo  {journal} {Medical and Biological
  Engineering and Computing}\ } (\bibinfo {year} {2020}),\
  10.1007/s11517-019-02092-w}\BibitemShut {NoStop}%
\bibitem [{\citenamefont {Borojeni}\ \emph {et~al.}(2020)\citenamefont
  {Borojeni}, \citenamefont {Garcia}, \citenamefont {Moghaddam}, \citenamefont
  {Frank-Ito}, \citenamefont {Kimbell}, \citenamefont {Laud}, \citenamefont
  {Koenig},\ and\ \citenamefont {Rhee}}]{Borojeni2020}%
  \BibitemOpen
  \bibfield  {author} {\bibinfo {author} {\bibfnamefont {A.~A.}\ \bibnamefont
  {Borojeni}}, \bibinfo {author} {\bibfnamefont {G.~J.}\ \bibnamefont
  {Garcia}}, \bibinfo {author} {\bibfnamefont {M.~G.}\ \bibnamefont
  {Moghaddam}}, \bibinfo {author} {\bibfnamefont {D.~O.}\ \bibnamefont
  {Frank-Ito}}, \bibinfo {author} {\bibfnamefont {J.~S.}\ \bibnamefont
  {Kimbell}}, \bibinfo {author} {\bibfnamefont {P.~W.}\ \bibnamefont {Laud}},
  \bibinfo {author} {\bibfnamefont {L.~J.}\ \bibnamefont {Koenig}}, \ and\
  \bibinfo {author} {\bibfnamefont {J.~S.}\ \bibnamefont {Rhee}},\ }\bibfield
  {title} {\enquote {\bibinfo {title} {{Normative ranges of nasal airflow
  variables in healthy adults}},}\ }\href {\doibase 10.1007/s11548-019-02023-y}
  {\bibfield  {journal} {\bibinfo  {journal} {International Journal of Computer
  Assisted Radiology and Surgery}\ } (\bibinfo {year} {2020}),\
  10.1007/s11548-019-02023-y}\BibitemShut {NoStop}%
\bibitem [{\citenamefont {Frank-Ito}\ \emph {et~al.}(2019)\citenamefont
  {Frank-Ito}, \citenamefont {Kimbell}, \citenamefont {Borojeni}, \citenamefont
  {Garcia},\ and\ \citenamefont {Rhee}}]{FrankIto2019}%
  \BibitemOpen
  \bibfield  {author} {\bibinfo {author} {\bibfnamefont {D.~O.}\ \bibnamefont
  {Frank-Ito}}, \bibinfo {author} {\bibfnamefont {J.~S.}\ \bibnamefont
  {Kimbell}}, \bibinfo {author} {\bibfnamefont {A.~A.}\ \bibnamefont
  {Borojeni}}, \bibinfo {author} {\bibfnamefont {G.~J.}\ \bibnamefont
  {Garcia}}, \ and\ \bibinfo {author} {\bibfnamefont {J.~S.}\ \bibnamefont
  {Rhee}},\ }\bibfield  {title} {\enquote {\bibinfo {title} {{A hierarchical
  stepwise approach to evaluate nasal patency after virtual surgery for nasal
  airway obstruction}},}\ }\href {\doibase 10.1016/j.clinbiomech.2018.12.014}
  {\bibfield  {journal} {\bibinfo  {journal} {Clinical Biomechanics}\ }
  (\bibinfo {year} {2019}),\ 10.1016/j.clinbiomech.2018.12.014}\BibitemShut
  {NoStop}%
\bibitem [{\citenamefont {Radulesco}\ \emph {et~al.}(2019)\citenamefont
  {Radulesco}, \citenamefont {Meister}, \citenamefont {Bouchet}, \citenamefont
  {Giordano}, \citenamefont {Dessi}, \citenamefont {Perrier},\ and\
  \citenamefont {Michel}}]{Radulesco2019}%
  \BibitemOpen
  \bibfield  {author} {\bibinfo {author} {\bibfnamefont {T.}~\bibnamefont
  {Radulesco}}, \bibinfo {author} {\bibfnamefont {L.}~\bibnamefont {Meister}},
  \bibinfo {author} {\bibfnamefont {G.}~\bibnamefont {Bouchet}}, \bibinfo
  {author} {\bibfnamefont {J.}~\bibnamefont {Giordano}}, \bibinfo {author}
  {\bibfnamefont {P.}~\bibnamefont {Dessi}}, \bibinfo {author} {\bibfnamefont
  {P.}~\bibnamefont {Perrier}}, \ and\ \bibinfo {author} {\bibfnamefont
  {J.}~\bibnamefont {Michel}},\ }\bibfield  {title} {\enquote {\bibinfo {title}
  {{Functional relevance of computational fluid dynamics in the field of nasal
  obstruction: A literature review}},}\ }\href {\doibase 10.1111/coa.13396}
  {\bibfield  {journal} {\bibinfo  {journal} {Clinical Otolaryngology}\ }
  (\bibinfo {year} {2019}),\ 10.1111/coa.13396}\BibitemShut {NoStop}%
\bibitem [{\citenamefont {Huang}\ \emph {et~al.}(2019)\citenamefont {Huang},
  \citenamefont {Nedanoski}, \citenamefont {Fletcher}, \citenamefont {Singh},
  \citenamefont {Schmid}, \citenamefont {Young}, \citenamefont {Stow},
  \citenamefont {Bi}, \citenamefont {Traini}, \citenamefont {Wong},
  \citenamefont {Phillips}, \citenamefont {Grunstein},\ and\ \citenamefont
  {Kim}}]{Huang2019}%
  \BibitemOpen
  \bibfield  {author} {\bibinfo {author} {\bibfnamefont {R.}~\bibnamefont
  {Huang}}, \bibinfo {author} {\bibfnamefont {A.}~\bibnamefont {Nedanoski}},
  \bibinfo {author} {\bibfnamefont {D.~F.}\ \bibnamefont {Fletcher}}, \bibinfo
  {author} {\bibfnamefont {N.}~\bibnamefont {Singh}}, \bibinfo {author}
  {\bibfnamefont {J.}~\bibnamefont {Schmid}}, \bibinfo {author} {\bibfnamefont
  {P.~M.}\ \bibnamefont {Young}}, \bibinfo {author} {\bibfnamefont
  {N.}~\bibnamefont {Stow}}, \bibinfo {author} {\bibfnamefont {L.}~\bibnamefont
  {Bi}}, \bibinfo {author} {\bibfnamefont {D.}~\bibnamefont {Traini}}, \bibinfo
  {author} {\bibfnamefont {E.}~\bibnamefont {Wong}}, \bibinfo {author}
  {\bibfnamefont {C.~L.}\ \bibnamefont {Phillips}}, \bibinfo {author}
  {\bibfnamefont {R.~R.}\ \bibnamefont {Grunstein}}, \ and\ \bibinfo {author}
  {\bibfnamefont {J.}~\bibnamefont {Kim}},\ }\bibfield  {title} {\enquote
  {\bibinfo {title} {{An automated segmentation framework for nasal
  computational fluid dynamics analysis in computed tomography}},}\ }\href
  {\doibase 10.1016/j.compbiomed.2019.103505} {\bibfield  {journal} {\bibinfo
  {journal} {Computers in Biology and Medicine}\ } (\bibinfo {year} {2019}),\
  10.1016/j.compbiomed.2019.103505}\BibitemShut {NoStop}%
\bibitem [{\citenamefont {Li}\ \emph {et~al.}(2017)\citenamefont {Li},
  \citenamefont {Jiang}, \citenamefont {Dong},\ and\ \citenamefont
  {Zhao}}]{Li2017}%
  \BibitemOpen
  \bibfield  {author} {\bibinfo {author} {\bibfnamefont {C.}~\bibnamefont
  {Li}}, \bibinfo {author} {\bibfnamefont {J.}~\bibnamefont {Jiang}}, \bibinfo
  {author} {\bibfnamefont {H.}~\bibnamefont {Dong}}, \ and\ \bibinfo {author}
  {\bibfnamefont {K.}~\bibnamefont {Zhao}},\ }\bibfield  {title} {\enquote
  {\bibinfo {title} {{Computational modeling and validation of human nasal
  airflow under various breathing conditions}},}\ }\href {\doibase
  10.1016/j.jbiomech.2017.08.031} {\bibfield  {journal} {\bibinfo  {journal}
  {Journal of Biomechanics}\ } (\bibinfo {year} {2017}),\
  10.1016/j.jbiomech.2017.08.031}\BibitemShut {NoStop}%
\bibitem [{\citenamefont {Zhao}\ \emph {et~al.}(2004)\citenamefont {Zhao},
  \citenamefont {Scherer}, \citenamefont {Hajiloo},\ and\ \citenamefont
  {Dalton}}]{Zhao2004}%
  \BibitemOpen
  \bibfield  {author} {\bibinfo {author} {\bibfnamefont {K.}~\bibnamefont
  {Zhao}}, \bibinfo {author} {\bibfnamefont {P.~W.}\ \bibnamefont {Scherer}},
  \bibinfo {author} {\bibfnamefont {S.~A.}\ \bibnamefont {Hajiloo}}, \ and\
  \bibinfo {author} {\bibfnamefont {P.}~\bibnamefont {Dalton}},\ }\bibfield
  {title} {\enquote {\bibinfo {title} {{Effect of anatomy on human nasal air
  flow and odorant transport patterns: Implications for olfaction}},}\ }\href
  {\doibase 10.1093/chemse/bjh033} {\bibfield  {journal} {\bibinfo  {journal}
  {Chemical Senses}\ } (\bibinfo {year} {2004}),\
  10.1093/chemse/bjh033}\BibitemShut {NoStop}%
\bibitem [{\citenamefont {Shang}, \citenamefont {Inthavong},\ and\
  \citenamefont {Tu}(2015)}]{Shang2015a}%
  \BibitemOpen
  \bibfield  {author} {\bibinfo {author} {\bibfnamefont {Y.~D.}\ \bibnamefont
  {Shang}}, \bibinfo {author} {\bibfnamefont {K.}~\bibnamefont {Inthavong}}, \
  and\ \bibinfo {author} {\bibfnamefont {J.~Y.}\ \bibnamefont {Tu}},\
  }\bibfield  {title} {\enquote {\bibinfo {title} {{Detailed micro-particle
  deposition patterns in the human nasal cavity influenced by the breathing
  zone}},}\ }\href {\doibase 10.1016/j.compfluid.2015.02.020} {\bibfield
  {journal} {\bibinfo  {journal} {Computers and Fluids}\ } (\bibinfo {year}
  {2015}),\ 10.1016/j.compfluid.2015.02.020}\BibitemShut {NoStop}%
\bibitem [{\citenamefont {Garcia}\ \emph {et~al.}(2007)\citenamefont {Garcia},
  \citenamefont {Bailie}, \citenamefont {Martins},\ and\ \citenamefont
  {Kimbell}}]{Garcia2007}%
  \BibitemOpen
  \bibfield  {author} {\bibinfo {author} {\bibfnamefont {G.~J.}\ \bibnamefont
  {Garcia}}, \bibinfo {author} {\bibfnamefont {N.}~\bibnamefont {Bailie}},
  \bibinfo {author} {\bibfnamefont {D.~A.}\ \bibnamefont {Martins}}, \ and\
  \bibinfo {author} {\bibfnamefont {J.~S.}\ \bibnamefont {Kimbell}},\
  }\bibfield  {title} {\enquote {\bibinfo {title} {{Atrophic rhinitis: A CFD
  study of air conditioning in the nasal cavity}},}\ }\href {\doibase
  10.1152/japplphysiol.01118.2006} {\bibfield  {journal} {\bibinfo  {journal}
  {Journal of Applied Physiology}\ } (\bibinfo {year} {2007}),\
  10.1152/japplphysiol.01118.2006}\BibitemShut {NoStop}%
\bibitem [{\citenamefont {Ge}, \citenamefont {Inthavong},\ and\ \citenamefont
  {Tu}(2012)}]{Ge2012}%
  \BibitemOpen
  \bibfield  {author} {\bibinfo {author} {\bibfnamefont {Q.~J.}\ \bibnamefont
  {Ge}}, \bibinfo {author} {\bibfnamefont {K.}~\bibnamefont {Inthavong}}, \
  and\ \bibinfo {author} {\bibfnamefont {J.~Y.}\ \bibnamefont {Tu}},\
  }\bibfield  {title} {\enquote {\bibinfo {title} {{Local deposition fractions
  of ultrafine particles in a human nasal-sinus cavity CFD model}},}\ }\href
  {\doibase 10.3109/08958378.2012.694494} {\bibfield  {journal} {\bibinfo
  {journal} {Inhalation Toxicology}\ } (\bibinfo {year} {2012}),\
  10.3109/08958378.2012.694494}\BibitemShut {NoStop}%
\bibitem [{\citenamefont {Goodarzi-Ardakani}\ \emph {et~al.}(2016)\citenamefont
  {Goodarzi-Ardakani}, \citenamefont {Taeibi-Rahni}, \citenamefont {Salimi},\
  and\ \citenamefont {Ahmadi}}]{Goodarzi-Ardakani2016}%
  \BibitemOpen
  \bibfield  {author} {\bibinfo {author} {\bibfnamefont {V.}~\bibnamefont
  {Goodarzi-Ardakani}}, \bibinfo {author} {\bibfnamefont {M.}~\bibnamefont
  {Taeibi-Rahni}}, \bibinfo {author} {\bibfnamefont {M.~R.}\ \bibnamefont
  {Salimi}}, \ and\ \bibinfo {author} {\bibfnamefont {G.}~\bibnamefont
  {Ahmadi}},\ }\bibfield  {title} {\enquote {\bibinfo {title} {{Computational
  simulation of temperature and velocity distribution in human upper
  respiratory airway during inhalation of hot air}},}\ }\href {\doibase
  10.1016/j.resp.2016.01.001} {\bibfield  {journal} {\bibinfo  {journal}
  {Respiratory Physiology and Neurobiology}\ } (\bibinfo {year} {2016}),\
  10.1016/j.resp.2016.01.001}\BibitemShut {NoStop}%
\bibitem [{\citenamefont {Lindemann}\ \emph {et~al.}(2005)\citenamefont
  {Lindemann}, \citenamefont {Brambs}, \citenamefont {Keck}, \citenamefont
  {Wiesmiller}, \citenamefont {Rettinger},\ and\ \citenamefont
  {Pless}}]{Lindemann2005}%
  \BibitemOpen
  \bibfield  {author} {\bibinfo {author} {\bibfnamefont {J.}~\bibnamefont
  {Lindemann}}, \bibinfo {author} {\bibfnamefont {H.~J.}\ \bibnamefont
  {Brambs}}, \bibinfo {author} {\bibfnamefont {T.}~\bibnamefont {Keck}},
  \bibinfo {author} {\bibfnamefont {K.~M.}\ \bibnamefont {Wiesmiller}},
  \bibinfo {author} {\bibfnamefont {G.}~\bibnamefont {Rettinger}}, \ and\
  \bibinfo {author} {\bibfnamefont {D.}~\bibnamefont {Pless}},\ }\bibfield
  {title} {\enquote {\bibinfo {title} {{Numerical simulation of intranasal
  airflow after radical sinus surgery}},}\ }\href {\doibase
  10.1016/j.amjoto.2005.02.010} {\bibfield  {journal} {\bibinfo  {journal}
  {American Journal of Otolaryngology - Head and Neck Medicine and Surgery}\ }
  (\bibinfo {year} {2005}),\ 10.1016/j.amjoto.2005.02.010}\BibitemShut
  {NoStop}%
\bibitem [{\citenamefont {H{\"{o}}rschler}\ \emph {et~al.}(2006)\citenamefont
  {H{\"{o}}rschler}, \citenamefont {Br{\"{u}}cker}, \citenamefont
  {Schr{\"{o}}der},\ and\ \citenamefont {Meinke}}]{Horschler2006}%
  \BibitemOpen
  \bibfield  {author} {\bibinfo {author} {\bibfnamefont {I.}~\bibnamefont
  {H{\"{o}}rschler}}, \bibinfo {author} {\bibfnamefont {C.}~\bibnamefont
  {Br{\"{u}}cker}}, \bibinfo {author} {\bibfnamefont {W.}~\bibnamefont
  {Schr{\"{o}}der}}, \ and\ \bibinfo {author} {\bibfnamefont {M.}~\bibnamefont
  {Meinke}},\ }\bibfield  {title} {\enquote {\bibinfo {title} {{Investigation
  of the impact of the geometry on the nose flow}},}\ }\href {\doibase
  10.1016/j.euromechflu.2005.11.006} {\bibfield  {journal} {\bibinfo  {journal}
  {European Journal of Mechanics, B/Fluids}\ } (\bibinfo {year} {2006}),\
  10.1016/j.euromechflu.2005.11.006}\BibitemShut {NoStop}%
\bibitem [{\citenamefont {Wen}\ \emph {et~al.}(2008)\citenamefont {Wen},
  \citenamefont {Inthavong}, \citenamefont {Tu},\ and\ \citenamefont
  {Wang}}]{Wen2008}%
  \BibitemOpen
  \bibfield  {author} {\bibinfo {author} {\bibfnamefont {J.}~\bibnamefont
  {Wen}}, \bibinfo {author} {\bibfnamefont {K.}~\bibnamefont {Inthavong}},
  \bibinfo {author} {\bibfnamefont {J.}~\bibnamefont {Tu}}, \ and\ \bibinfo
  {author} {\bibfnamefont {S.}~\bibnamefont {Wang}},\ }\bibfield  {title}
  {\enquote {\bibinfo {title} {{Numerical simulations for detailed airflow
  dynamics in a human nasal cavity}},}\ }\href {\doibase
  10.1016/j.resp.2008.01.012} {\bibfield  {journal} {\bibinfo  {journal}
  {Respiratory Physiology and Neurobiology}\ } (\bibinfo {year} {2008}),\
  10.1016/j.resp.2008.01.012}\BibitemShut {NoStop}%
\bibitem [{\citenamefont {Inthavong}, \citenamefont {Tu},\ and\ \citenamefont
  {Heschl}(2011)}]{Inthavong2011}%
  \BibitemOpen
  \bibfield  {author} {\bibinfo {author} {\bibfnamefont {K.}~\bibnamefont
  {Inthavong}}, \bibinfo {author} {\bibfnamefont {J.}~\bibnamefont {Tu}}, \
  and\ \bibinfo {author} {\bibfnamefont {C.}~\bibnamefont {Heschl}},\
  }\bibfield  {title} {\enquote {\bibinfo {title} {{Micron particle deposition
  in the nasal cavity using the v2-f model}},}\ }\href {\doibase
  10.1016/j.compfluid.2011.08.013} {\bibfield  {journal} {\bibinfo  {journal}
  {Computers and Fluids}\ } (\bibinfo {year} {2011}),\
  10.1016/j.compfluid.2011.08.013}\BibitemShut {NoStop}%
\bibitem [{\citenamefont {Kleinstreuer}\ and\ \citenamefont
  {Zhang}(2003)}]{Kleinstreuer2003}%
  \BibitemOpen
  \bibfield  {author} {\bibinfo {author} {\bibfnamefont {C.}~\bibnamefont
  {Kleinstreuer}}\ and\ \bibinfo {author} {\bibfnamefont {Z.}~\bibnamefont
  {Zhang}},\ }\bibfield  {title} {\enquote {\bibinfo {title}
  {{Laminar-to-turbulent fluid-particle flows in a human airway model}},}\
  }\href {\doibase 10.1016/S0301-9322(02)00131-3} {\bibfield  {journal}
  {\bibinfo  {journal} {International Journal of Multiphase Flow}\ } (\bibinfo
  {year} {2003}),\ 10.1016/S0301-9322(02)00131-3}\BibitemShut {NoStop}%
\bibitem [{\citenamefont {Zhu}\ \emph {et~al.}(2013)\citenamefont {Zhu},
  \citenamefont {Lee}, \citenamefont {Lim}, \citenamefont {Lee}, \citenamefont
  {{Teo Li San}},\ and\ \citenamefont {Wang}}]{Zhu2013}%
  \BibitemOpen
  \bibfield  {author} {\bibinfo {author} {\bibfnamefont {J.~H.}\ \bibnamefont
  {Zhu}}, \bibinfo {author} {\bibfnamefont {H.~P.}\ \bibnamefont {Lee}},
  \bibinfo {author} {\bibfnamefont {K.~M.}\ \bibnamefont {Lim}}, \bibinfo
  {author} {\bibfnamefont {S.~J.}\ \bibnamefont {Lee}}, \bibinfo {author}
  {\bibfnamefont {L.}~\bibnamefont {{Teo Li San}}}, \ and\ \bibinfo {author}
  {\bibfnamefont {D.~Y.}\ \bibnamefont {Wang}},\ }\bibfield  {title} {\enquote
  {\bibinfo {title} {{Inspirational airflow patterns in deviated noses: A
  numerical study}},}\ }\href {\doibase 10.1080/10255842.2012.670850}
  {\bibfield  {journal} {\bibinfo  {journal} {Computer Methods in Biomechanics
  and Biomedical Engineering}\ } (\bibinfo {year} {2013}),\
  10.1080/10255842.2012.670850}\BibitemShut {NoStop}%
\bibitem [{\citenamefont {Kim}, \citenamefont {Xi},\ and\ \citenamefont
  {Si}(2013)}]{Kim2013a}%
  \BibitemOpen
  \bibfield  {author} {\bibinfo {author} {\bibfnamefont {J.~W.}\ \bibnamefont
  {Kim}}, \bibinfo {author} {\bibfnamefont {J.}~\bibnamefont {Xi}}, \ and\
  \bibinfo {author} {\bibfnamefont {X.~A.}\ \bibnamefont {Si}},\ }\bibfield
  {title} {\enquote {\bibinfo {title} {{Dynamic growth and deposition of
  hygroscopic aerosols in the nasal airway of a 5-year-old child}},}\ }\href
  {\doibase 10.1002/cnm.2490} {\bibfield  {journal} {\bibinfo  {journal}
  {International Journal for Numerical Methods in Biomedical Engineering}\ }
  (\bibinfo {year} {2013}),\ 10.1002/cnm.2490}\BibitemShut {NoStop}%
\bibitem [{\citenamefont {Ito}\ \emph {et~al.}(2017)\citenamefont {Ito},
  \citenamefont {Mitsumune}, \citenamefont {Kuga}, \citenamefont {Phuong},
  \citenamefont {Tani},\ and\ \citenamefont {Inthavong}}]{Ito2017}%
  \BibitemOpen
  \bibfield  {author} {\bibinfo {author} {\bibfnamefont {K.}~\bibnamefont
  {Ito}}, \bibinfo {author} {\bibfnamefont {K.}~\bibnamefont {Mitsumune}},
  \bibinfo {author} {\bibfnamefont {K.}~\bibnamefont {Kuga}}, \bibinfo {author}
  {\bibfnamefont {N.~L.}\ \bibnamefont {Phuong}}, \bibinfo {author}
  {\bibfnamefont {K.}~\bibnamefont {Tani}}, \ and\ \bibinfo {author}
  {\bibfnamefont {K.}~\bibnamefont {Inthavong}},\ }\bibfield  {title} {\enquote
  {\bibinfo {title} {{Prediction of convective heat transfer coefficients for
  the upper respiratory tracts of rat, dog, monkey, and humans}},}\ }\href
  {\doibase 10.1177/1420326X16662111} {\bibfield  {journal} {\bibinfo
  {journal} {Indoor and Built Environment}\ } (\bibinfo {year} {2017}),\
  10.1177/1420326X16662111}\BibitemShut {NoStop}%
\bibitem [{\citenamefont {Kim}\ \emph {et~al.}(2013)\citenamefont {Kim},
  \citenamefont {Na}, \citenamefont {Kim},\ and\ \citenamefont
  {Chung}}]{Kim2013}%
  \BibitemOpen
  \bibfield  {author} {\bibinfo {author} {\bibfnamefont {S.~K.}\ \bibnamefont
  {Kim}}, \bibinfo {author} {\bibfnamefont {Y.}~\bibnamefont {Na}}, \bibinfo
  {author} {\bibfnamefont {J.~I.}\ \bibnamefont {Kim}}, \ and\ \bibinfo
  {author} {\bibfnamefont {S.~K.}\ \bibnamefont {Chung}},\ }\bibfield  {title}
  {\enquote {\bibinfo {title} {{Patient specific CFD models of nasal airflow:
  Overview of methods and challenges}},}\ }\href {\doibase
  10.1016/j.jbiomech.2012.11.022} {\bibfield  {journal} {\bibinfo  {journal}
  {Journal of Biomechanics}\ } (\bibinfo {year} {2013}),\
  10.1016/j.jbiomech.2012.11.022}\BibitemShut {NoStop}%
\bibitem [{\citenamefont {Wang}\ \emph {et~al.}(2016)\citenamefont {Wang},
  \citenamefont {Chen}, \citenamefont {Wang}, \citenamefont {Chen},\ and\
  \citenamefont {Deng}}]{Wang2016}%
  \BibitemOpen
  \bibfield  {author} {\bibinfo {author} {\bibfnamefont {T.}~\bibnamefont
  {Wang}}, \bibinfo {author} {\bibfnamefont {D.}~\bibnamefont {Chen}}, \bibinfo
  {author} {\bibfnamefont {P.~H.}\ \bibnamefont {Wang}}, \bibinfo {author}
  {\bibfnamefont {J.}~\bibnamefont {Chen}}, \ and\ \bibinfo {author}
  {\bibfnamefont {J.}~\bibnamefont {Deng}},\ }\bibfield  {title} {\enquote
  {\bibinfo {title} {{Investigation on the nasal airflow characteristics of
  anterior nasal cavity stenosis}},}\ }\href {\doibase
  10.1590/1414-431X20165182} {\bibfield  {journal} {\bibinfo  {journal}
  {Brazilian Journal of Medical and Biological Research}\ } (\bibinfo {year}
  {2016}),\ 10.1590/1414-431X20165182}\BibitemShut {NoStop}%
\bibitem [{\citenamefont {Wakayama}, \citenamefont {Suzuki},\ and\
  \citenamefont {Tanuma}(2016)}]{Wakayama2016}%
  \BibitemOpen
  \bibfield  {author} {\bibinfo {author} {\bibfnamefont {T.}~\bibnamefont
  {Wakayama}}, \bibinfo {author} {\bibfnamefont {M.}~\bibnamefont {Suzuki}}, \
  and\ \bibinfo {author} {\bibfnamefont {T.}~\bibnamefont {Tanuma}},\
  }\bibfield  {title} {\enquote {\bibinfo {title} {{Effect of nasal obstruction
  on continuous positive airway pressure treatment: Computational fluid
  dynamics analyses}},}\ }\href {\doibase 10.1371/journal.pone.0150951}
  {\bibfield  {journal} {\bibinfo  {journal} {PLoS ONE}\ } (\bibinfo {year}
  {2016}),\ 10.1371/journal.pone.0150951}\BibitemShut {NoStop}%
\bibitem [{\citenamefont {Kim}\ \emph {et~al.}(2014)\citenamefont {Kim},
  \citenamefont {Heo}, \citenamefont {Seo}, \citenamefont {Na},\ and\
  \citenamefont {Chung}}]{Kim2014}%
  \BibitemOpen
  \bibfield  {author} {\bibinfo {author} {\bibfnamefont {S.~K.}\ \bibnamefont
  {Kim}}, \bibinfo {author} {\bibfnamefont {G.~E.}\ \bibnamefont {Heo}},
  \bibinfo {author} {\bibfnamefont {A.}~\bibnamefont {Seo}}, \bibinfo {author}
  {\bibfnamefont {Y.}~\bibnamefont {Na}}, \ and\ \bibinfo {author}
  {\bibfnamefont {S.~K.}\ \bibnamefont {Chung}},\ }\bibfield  {title} {\enquote
  {\bibinfo {title} {{Correlation between nasal airflow characteristics and
  clinical relevance of nasal septal deviation to nasal airway obstruction}},}\
  }\href {\doibase 10.1016/j.resp.2013.12.010} {\bibfield  {journal} {\bibinfo
  {journal} {Respiratory Physiology and Neurobiology}\ } (\bibinfo {year}
  {2014}),\ 10.1016/j.resp.2013.12.010}\BibitemShut {NoStop}%
\bibitem [{\citenamefont {Dong}\ \emph {et~al.}(2018)\citenamefont {Dong},
  \citenamefont {Ma}, \citenamefont {Shang}, \citenamefont {Inthavong},
  \citenamefont {Qiu}, \citenamefont {Tu},\ and\ \citenamefont
  {Frank-Ito}}]{Dong2018}%
  \BibitemOpen
  \bibfield  {author} {\bibinfo {author} {\bibfnamefont {J.}~\bibnamefont
  {Dong}}, \bibinfo {author} {\bibfnamefont {J.}~\bibnamefont {Ma}}, \bibinfo
  {author} {\bibfnamefont {Y.}~\bibnamefont {Shang}}, \bibinfo {author}
  {\bibfnamefont {K.}~\bibnamefont {Inthavong}}, \bibinfo {author}
  {\bibfnamefont {D.}~\bibnamefont {Qiu}}, \bibinfo {author} {\bibfnamefont
  {J.}~\bibnamefont {Tu}}, \ and\ \bibinfo {author} {\bibfnamefont
  {D.}~\bibnamefont {Frank-Ito}},\ }\bibfield  {title} {\enquote {\bibinfo
  {title} {{Detailed nanoparticle exposure analysis among human nasal cavities
  with distinct vestibule phenotypes}},}\ }\href {\doibase
  10.1016/j.jaerosci.2018.05.001} {\bibfield  {journal} {\bibinfo  {journal}
  {Journal of Aerosol Science}\ } (\bibinfo {year} {2018}),\
  10.1016/j.jaerosci.2018.05.001}\BibitemShut {NoStop}%
\bibitem [{\citenamefont {Inthavong}\ \emph {et~al.}(2019)\citenamefont
  {Inthavong}, \citenamefont {Das}, \citenamefont {Singh},\ and\ \citenamefont
  {Sznitman}}]{Inthavong2019}%
  \BibitemOpen
  \bibfield  {author} {\bibinfo {author} {\bibfnamefont {K.}~\bibnamefont
  {Inthavong}}, \bibinfo {author} {\bibfnamefont {P.}~\bibnamefont {Das}},
  \bibinfo {author} {\bibfnamefont {N.}~\bibnamefont {Singh}}, \ and\ \bibinfo
  {author} {\bibfnamefont {J.}~\bibnamefont {Sznitman}},\ }\bibfield  {title}
  {\enquote {\bibinfo {title} {In silico approaches to respiratory nasal flows:
  A review},}\ }\href {\doibase https://doi.org/10.1016/j.jbiomech.2019.109434}
  {\bibfield  {journal} {\bibinfo  {journal} {Journal of Biomechanics}\
  }\textbf {\bibinfo {volume} {97}},\ \bibinfo {pages} {109434} (\bibinfo
  {year} {2019})}\BibitemShut {NoStop}%
\bibitem [{\citenamefont {Calmet}\ \emph {et~al.}(2016)\citenamefont {Calmet},
  \citenamefont {Gambaruto}, \citenamefont {Bates}, \citenamefont {Vazquez},
  \citenamefont {Houzeaux},\ and\ \citenamefont {Doorly}}]{Calmet2016}%
  \BibitemOpen
  \bibfield  {author} {\bibinfo {author} {\bibfnamefont {H.}~\bibnamefont
  {Calmet}}, \bibinfo {author} {\bibfnamefont {A.~M.}\ \bibnamefont
  {Gambaruto}}, \bibinfo {author} {\bibfnamefont {A.~J.}\ \bibnamefont
  {Bates}}, \bibinfo {author} {\bibfnamefont {M.}~\bibnamefont {Vazquez}},
  \bibinfo {author} {\bibfnamefont {G.}~\bibnamefont {Houzeaux}}, \ and\
  \bibinfo {author} {\bibfnamefont {D.~J.}\ \bibnamefont {Doorly}},\ }\bibfield
   {title} {\enquote {\bibinfo {title} {{Large-scale CFD simulations of the
  transitional and turbulent regime for the large human airways during rapid
  inhalation}},}\ }\href {\doibase 10.1016/j.compbiomed.2015.12.003} {\bibfield
   {journal} {\bibinfo  {journal} {Computers in Biology and Medicine}\ }
  (\bibinfo {year} {2016}),\ 10.1016/j.compbiomed.2015.12.003}\BibitemShut
  {NoStop}%
\bibitem [{\citenamefont {Lee}\ \emph {et~al.}(2010)\citenamefont {Lee},
  \citenamefont {Na}, \citenamefont {Kim},\ and\ \citenamefont
  {Chung}}]{Lee2010}%
  \BibitemOpen
  \bibfield  {author} {\bibinfo {author} {\bibfnamefont {J.~H.}\ \bibnamefont
  {Lee}}, \bibinfo {author} {\bibfnamefont {Y.}~\bibnamefont {Na}}, \bibinfo
  {author} {\bibfnamefont {S.~K.}\ \bibnamefont {Kim}}, \ and\ \bibinfo
  {author} {\bibfnamefont {S.~K.}\ \bibnamefont {Chung}},\ }\bibfield  {title}
  {\enquote {\bibinfo {title} {{Unsteady flow characteristics through a human
  nasal airway}},}\ }\href {\doibase 10.1016/j.resp.2010.05.010} {\bibfield
  {journal} {\bibinfo  {journal} {Respiratory Physiology and Neurobiology}\ }
  (\bibinfo {year} {2010}),\ 10.1016/j.resp.2010.05.010}\BibitemShut {NoStop}%
\bibitem [{\citenamefont {Liu}\ \emph {et~al.}(2007)\citenamefont {Liu},
  \citenamefont {Matida}, \citenamefont {Gu},\ and\ \citenamefont
  {Johnson}}]{Liu2007}%
  \BibitemOpen
  \bibfield  {author} {\bibinfo {author} {\bibfnamefont {Y.}~\bibnamefont
  {Liu}}, \bibinfo {author} {\bibfnamefont {E.~A.}\ \bibnamefont {Matida}},
  \bibinfo {author} {\bibfnamefont {J.}~\bibnamefont {Gu}}, \ and\ \bibinfo
  {author} {\bibfnamefont {M.~R.}\ \bibnamefont {Johnson}},\ }\bibfield
  {title} {\enquote {\bibinfo {title} {{Numerical simulation of aerosol
  deposition in a 3-D human nasal cavity using RANS, RANS/EIM, and LES}},}\
  }\href {\doibase 10.1016/j.jaerosci.2007.05.003} {\bibfield  {journal}
  {\bibinfo  {journal} {Journal of Aerosol Science}\ } (\bibinfo {year}
  {2007}),\ 10.1016/j.jaerosci.2007.05.003}\BibitemShut {NoStop}%
\bibitem [{\citenamefont {Ghahramani}\ \emph {et~al.}(2017)\citenamefont
  {Ghahramani}, \citenamefont {Abouali}, \citenamefont {Emdad},\ and\
  \citenamefont {Ahmadi}}]{Ghahramani2017}%
  \BibitemOpen
  \bibfield  {author} {\bibinfo {author} {\bibfnamefont {E.}~\bibnamefont
  {Ghahramani}}, \bibinfo {author} {\bibfnamefont {O.}~\bibnamefont {Abouali}},
  \bibinfo {author} {\bibfnamefont {H.}~\bibnamefont {Emdad}}, \ and\ \bibinfo
  {author} {\bibfnamefont {G.}~\bibnamefont {Ahmadi}},\ }\bibfield  {title}
  {\enquote {\bibinfo {title} {{Numerical investigation of turbulent airflow
  and microparticle deposition in a realistic model of human upper airway using
  LES}},}\ }\href {\doibase 10.1016/j.compfluid.2017.08.003} {\bibfield
  {journal} {\bibinfo  {journal} {Computers and Fluids}\ } (\bibinfo {year}
  {2017}),\ 10.1016/j.compfluid.2017.08.003}\BibitemShut {NoStop}%
\bibitem [{\citenamefont {Menter}(2018)}]{Menter2018}%
  \BibitemOpen
  \bibfield  {author} {\bibinfo {author} {\bibfnamefont {F.}~\bibnamefont
  {Menter}},\ }\bibfield  {title} {\enquote {\bibinfo {title} {{Stress-blended
  eddy simulation (SBES) - A new paradigm in hybrid RANS-LES modeling}},}\
  }\href {\doibase 10.1007/978-3-319-70031-13} {\bibfield  {journal} {\bibinfo
  {journal} {Notes on Numerical Fluid Mechanics and Multidisciplinary Design}\
  } (\bibinfo {year} {2018}),\ 10.1007/978-3-319-70031-13}\BibitemShut
  {NoStop}%
\bibitem [{\citenamefont {Brown}\ \emph {et~al.}(2020)\citenamefont {Brown},
  \citenamefont {Fletcher}, \citenamefont {Leggoe},\ and\ \citenamefont
  {Whyte}}]{Brown2020}%
  \BibitemOpen
  \bibfield  {author} {\bibinfo {author} {\bibfnamefont {G.~J.}\ \bibnamefont
  {Brown}}, \bibinfo {author} {\bibfnamefont {D.~F.}\ \bibnamefont {Fletcher}},
  \bibinfo {author} {\bibfnamefont {J.~W.}\ \bibnamefont {Leggoe}}, \ and\
  \bibinfo {author} {\bibfnamefont {D.~S.}\ \bibnamefont {Whyte}},\ }\bibfield
  {title} {\enquote {\bibinfo {title} {{Application of hybrid RANS-LES models
  to the prediction of flow behaviour in an industrial crystalliser}},}\ }\href
  {\doibase 10.1016/j.apm.2019.09.032} {\bibfield  {journal} {\bibinfo
  {journal} {Applied Mathematical Modelling}\ } (\bibinfo {year} {2020}),\
  10.1016/j.apm.2019.09.032}\BibitemShut {NoStop}%
\bibitem [{\citenamefont {Weinhold}\ and\ \citenamefont
  {Mlynski}(2004)}]{Weinhold2004}%
  \BibitemOpen
  \bibfield  {author} {\bibinfo {author} {\bibfnamefont {I.}~\bibnamefont
  {Weinhold}}\ and\ \bibinfo {author} {\bibfnamefont {G.}~\bibnamefont
  {Mlynski}},\ }\bibfield  {title} {\enquote {\bibinfo {title} {{Numerical
  simulation of airflow in the human nose}},}\ }\href {\doibase
  10.1007/s00405-003-0675-y} {\bibfield  {journal} {\bibinfo  {journal}
  {European Archives of Oto-Rhino-Laryngology}\ } (\bibinfo {year} {2004}),\
  10.1007/s00405-003-0675-y}\BibitemShut {NoStop}%
\bibitem [{\citenamefont {Inthavong}, \citenamefont {Shang},\ and\
  \citenamefont {Tu}(2014)}]{Inthavong2014}%
  \BibitemOpen
  \bibfield  {author} {\bibinfo {author} {\bibfnamefont {K.}~\bibnamefont
  {Inthavong}}, \bibinfo {author} {\bibfnamefont {Y.}~\bibnamefont {Shang}}, \
  and\ \bibinfo {author} {\bibfnamefont {J.}~\bibnamefont {Tu}},\ }\bibfield
  {title} {\enquote {\bibinfo {title} {{Surface mapping for visualization of
  wall stresses during inhalation in a human nasal cavity}},}\ }\href {\doibase
  10.1016/j.resp.2013.09.004} {\bibfield  {journal} {\bibinfo  {journal}
  {Respiratory Physiology and Neurobiology}\ } (\bibinfo {year} {2014}),\
  10.1016/j.resp.2013.09.004}\BibitemShut {NoStop}%
\bibitem [{\citenamefont {Shang}\ \emph {et~al.}(2015)\citenamefont {Shang},
  \citenamefont {Dong}, \citenamefont {Inthavong},\ and\ \citenamefont
  {Tu}}]{Shang2015}%
  \BibitemOpen
  \bibfield  {author} {\bibinfo {author} {\bibfnamefont {Y.}~\bibnamefont
  {Shang}}, \bibinfo {author} {\bibfnamefont {J.}~\bibnamefont {Dong}},
  \bibinfo {author} {\bibfnamefont {K.}~\bibnamefont {Inthavong}}, \ and\
  \bibinfo {author} {\bibfnamefont {J.}~\bibnamefont {Tu}},\ }\bibfield
  {title} {\enquote {\bibinfo {title} {{Comparative numerical modeling of
  inhaled micron-sized particle deposition in human and rat nasal cavities}},}\
  }\href {\doibase 10.3109/08958378.2015.1088600} {\bibfield  {journal}
  {\bibinfo  {journal} {Inhalation Toxicology}\ } (\bibinfo {year} {2015}),\
  10.3109/08958378.2015.1088600}\BibitemShut {NoStop}%
\bibitem [{\citenamefont {Dong}\ \emph {et~al.}(2016)\citenamefont {Dong},
  \citenamefont {Shang}, \citenamefont {Inthavong}, \citenamefont {Tu},
  \citenamefont {Chen}, \citenamefont {Bai}, \citenamefont {Wang},\ and\
  \citenamefont {Chen}}]{Dong2016}%
  \BibitemOpen
  \bibfield  {author} {\bibinfo {author} {\bibfnamefont {J.}~\bibnamefont
  {Dong}}, \bibinfo {author} {\bibfnamefont {Y.}~\bibnamefont {Shang}},
  \bibinfo {author} {\bibfnamefont {K.}~\bibnamefont {Inthavong}}, \bibinfo
  {author} {\bibfnamefont {J.}~\bibnamefont {Tu}}, \bibinfo {author}
  {\bibfnamefont {R.}~\bibnamefont {Chen}}, \bibinfo {author} {\bibfnamefont
  {R.}~\bibnamefont {Bai}}, \bibinfo {author} {\bibfnamefont {D.}~\bibnamefont
  {Wang}}, \ and\ \bibinfo {author} {\bibfnamefont {C.}~\bibnamefont {Chen}},\
  }\bibfield  {title} {\enquote {\bibinfo {title} {{Comparative numerical
  modeling of inhaled nanoparticle deposition in human and rat nasal
  cavities}},}\ }\href {\doibase 10.1093/toxsci/kfw087} {\bibfield  {journal}
  {\bibinfo  {journal} {Toxicological Sciences}\ } (\bibinfo {year} {2016}),\
  10.1093/toxsci/kfw087}\BibitemShut {NoStop}%
\bibitem [{\citenamefont {Croce}\ \emph {et~al.}(2006)\citenamefont {Croce},
  \citenamefont {Fodil}, \citenamefont {Durand}, \citenamefont {Sbirlea-Apiou},
  \citenamefont {Caillibotte}, \citenamefont {Papon}, \citenamefont {Blondeau},
  \citenamefont {Coste}, \citenamefont {Isabey},\ and\ \citenamefont
  {Louis}}]{Croce2006}%
  \BibitemOpen
  \bibfield  {author} {\bibinfo {author} {\bibfnamefont {C.}~\bibnamefont
  {Croce}}, \bibinfo {author} {\bibfnamefont {R.}~\bibnamefont {Fodil}},
  \bibinfo {author} {\bibfnamefont {M.}~\bibnamefont {Durand}}, \bibinfo
  {author} {\bibfnamefont {G.}~\bibnamefont {Sbirlea-Apiou}}, \bibinfo {author}
  {\bibfnamefont {G.}~\bibnamefont {Caillibotte}}, \bibinfo {author}
  {\bibfnamefont {J.~F.}\ \bibnamefont {Papon}}, \bibinfo {author}
  {\bibfnamefont {J.~R.}\ \bibnamefont {Blondeau}}, \bibinfo {author}
  {\bibfnamefont {A.}~\bibnamefont {Coste}}, \bibinfo {author} {\bibfnamefont
  {D.}~\bibnamefont {Isabey}}, \ and\ \bibinfo {author} {\bibfnamefont
  {B.}~\bibnamefont {Louis}},\ }\bibfield  {title} {\enquote {\bibinfo {title}
  {{In Vitro Experiments and Numerical Simulations of Airflow in Realistic
  Nasal Airway Geometry}},}\ }\href {\doibase 10.1007/s10439-006-9094-8}
  {\bibfield  {journal} {\bibinfo  {journal} {Annals of Biomedical
  Engineering}\ } (\bibinfo {year} {2006}),\
  10.1007/s10439-006-9094-8}\BibitemShut {NoStop}%
\bibitem [{\citenamefont {Zhao}\ \emph {et~al.}(2006)\citenamefont {Zhao},
  \citenamefont {Dalton}, \citenamefont {Yang},\ and\ \citenamefont
  {Scherer}}]{Zhao2006}%
  \BibitemOpen
  \bibfield  {author} {\bibinfo {author} {\bibfnamefont {K.}~\bibnamefont
  {Zhao}}, \bibinfo {author} {\bibfnamefont {P.}~\bibnamefont {Dalton}},
  \bibinfo {author} {\bibfnamefont {G.~C.}\ \bibnamefont {Yang}}, \ and\
  \bibinfo {author} {\bibfnamefont {P.~W.}\ \bibnamefont {Scherer}},\
  }\bibfield  {title} {\enquote {\bibinfo {title} {{Numerical modeling of
  turbulent and laminar airflow and odorant transport during sniffing in the
  human and rat nose}},}\ }\href {\doibase 10.1093/chemse/bjj008} {\bibfield
  {journal} {\bibinfo  {journal} {Chemical Senses}\ } (\bibinfo {year}
  {2006}),\ 10.1093/chemse/bjj008}\BibitemShut {NoStop}%
\bibitem [{\citenamefont {Segal}, \citenamefont {Kepler},\ and\ \citenamefont
  {Kimbell}(2008)}]{Segal2008}%
  \BibitemOpen
  \bibfield  {author} {\bibinfo {author} {\bibfnamefont {R.~A.}\ \bibnamefont
  {Segal}}, \bibinfo {author} {\bibfnamefont {G.~M.}\ \bibnamefont {Kepler}}, \
  and\ \bibinfo {author} {\bibfnamefont {J.~S.}\ \bibnamefont {Kimbell}},\
  }\bibfield  {title} {\enquote {\bibinfo {title} {{Effects of differences in
  nasal anatomy on airflow distribution: A comparison of four individuals at
  rest}},}\ }\href {\doibase 10.1007/s10439-008-9556-2} {\bibfield  {journal}
  {\bibinfo  {journal} {Annals of Biomedical Engineering}\ } (\bibinfo {year}
  {2008}),\ 10.1007/s10439-008-9556-2}\BibitemShut {NoStop}%
\bibitem [{\citenamefont {Schreck}\ \emph {et~al.}(1993)\citenamefont
  {Schreck}, \citenamefont {Sullivan}, \citenamefont {Ho},\ and\ \citenamefont
  {Chang}}]{Schreck1993}%
  \BibitemOpen
  \bibfield  {author} {\bibinfo {author} {\bibfnamefont {S.}~\bibnamefont
  {Schreck}}, \bibinfo {author} {\bibfnamefont {K.~J.}\ \bibnamefont
  {Sullivan}}, \bibinfo {author} {\bibfnamefont {C.~M.}\ \bibnamefont {Ho}}, \
  and\ \bibinfo {author} {\bibfnamefont {H.~K.}\ \bibnamefont {Chang}},\
  }\bibfield  {title} {\enquote {\bibinfo {title} {{Correlations between flow
  resistance and geometry in a model of the human nose}},}\ }\href {\doibase
  10.1152/jappl.1993.75.4.1767} {\bibfield  {journal} {\bibinfo  {journal}
  {Journal of Applied Physiology}\ } (\bibinfo {year} {1993}),\
  10.1152/jappl.1993.75.4.1767}\BibitemShut {NoStop}%
\bibitem [{\citenamefont {Zhu}\ \emph {et~al.}(2011)\citenamefont {Zhu},
  \citenamefont {Lee}, \citenamefont {Lim}, \citenamefont {Lee},\ and\
  \citenamefont {Wang}}]{Zhu2011}%
  \BibitemOpen
  \bibfield  {author} {\bibinfo {author} {\bibfnamefont {J.~H.}\ \bibnamefont
  {Zhu}}, \bibinfo {author} {\bibfnamefont {H.~P.}\ \bibnamefont {Lee}},
  \bibinfo {author} {\bibfnamefont {K.~M.}\ \bibnamefont {Lim}}, \bibinfo
  {author} {\bibfnamefont {S.~J.}\ \bibnamefont {Lee}}, \ and\ \bibinfo
  {author} {\bibfnamefont {D.~Y.}\ \bibnamefont {Wang}},\ }\bibfield  {title}
  {\enquote {\bibinfo {title} {Evaluation and comparison of nasal airway flow
  patterns among three subjects from caucasian, chinese and indian ethnic
  groups using computational fluid dynamics simulation},}\ }\href {\doibase
  10.1016/j.resp.2010.09.008} {\bibfield  {journal} {\bibinfo  {journal}
  {Respiratory Physiology and Neurobiology}\ }\textbf {\bibinfo {volume}
  {175}},\ \bibinfo {pages} {62--69} (\bibinfo {year} {2011})}\BibitemShut
  {NoStop}%
\bibitem [{\citenamefont {Shi}, \citenamefont {Kleinstreuer},\ and\
  \citenamefont {Zhang}(2006)}]{Shi2006}%
  \BibitemOpen
  \bibfield  {author} {\bibinfo {author} {\bibfnamefont {H.}~\bibnamefont
  {Shi}}, \bibinfo {author} {\bibfnamefont {C.}~\bibnamefont {Kleinstreuer}}, \
  and\ \bibinfo {author} {\bibfnamefont {Z.}~\bibnamefont {Zhang}},\ }\bibfield
   {title} {\enquote {\bibinfo {title} {{Laminar airflow and nanoparticle or
  vapor deposition in a human nasal cavity model}},}\ }\href {\doibase
  10.1115/1.2244574} {\bibfield  {journal} {\bibinfo  {journal} {Journal of
  Biomechanical Engineering}\ } (\bibinfo {year} {2006}),\
  10.1115/1.2244574}\BibitemShut {NoStop}%
\bibitem [{\citenamefont {Bruning}\ \emph {et~al.}(2020)\citenamefont
  {Bruning}, \citenamefont {Hildebrandt}, \citenamefont {Heppt}, \citenamefont
  {Schmidt}, \citenamefont {Lamecker}, \citenamefont {Szengel}, \citenamefont
  {Amiridze}, \citenamefont {Ramm}, \citenamefont {Bindernagel}, \citenamefont
  {Zachow},\ and\ \citenamefont {Goubergrits}}]{Bruning2020}%
  \BibitemOpen
  \bibfield  {author} {\bibinfo {author} {\bibfnamefont {J.}~\bibnamefont
  {Bruning}}, \bibinfo {author} {\bibfnamefont {T.}~\bibnamefont
  {Hildebrandt}}, \bibinfo {author} {\bibfnamefont {W.}~\bibnamefont {Heppt}},
  \bibinfo {author} {\bibfnamefont {N.}~\bibnamefont {Schmidt}}, \bibinfo
  {author} {\bibfnamefont {H.}~\bibnamefont {Lamecker}}, \bibinfo {author}
  {\bibfnamefont {A.}~\bibnamefont {Szengel}}, \bibinfo {author} {\bibfnamefont
  {N.}~\bibnamefont {Amiridze}}, \bibinfo {author} {\bibfnamefont
  {H.}~\bibnamefont {Ramm}}, \bibinfo {author} {\bibfnamefont {M.}~\bibnamefont
  {Bindernagel}}, \bibinfo {author} {\bibfnamefont {S.}~\bibnamefont {Zachow}},
  \ and\ \bibinfo {author} {\bibfnamefont {L.}~\bibnamefont {Goubergrits}},\
  }\bibfield  {title} {\enquote {\bibinfo {title} {{Characterization of the
  Airflow within an Average Geometry of the Healthy Human Nasal Cavity}},}\
  }\href {\doibase 10.1038/s41598-020-60755-3} {\bibfield  {journal} {\bibinfo
  {journal} {Scientific Reports}\ } (\bibinfo {year} {2020}),\
  10.1038/s41598-020-60755-3}\BibitemShut {NoStop}%
\bibitem [{\citenamefont {Sanmiguel-Rojas}\ \emph {et~al.}(2018)\citenamefont
  {Sanmiguel-Rojas}, \citenamefont {Burgos}, \citenamefont {del Pino},
  \citenamefont {Sevilla-Garcia},\ and\ \citenamefont
  {Esteban-Ortega}}]{Rojas2018}%
  \BibitemOpen
  \bibfield  {author} {\bibinfo {author} {\bibfnamefont {E.}~\bibnamefont
  {Sanmiguel-Rojas}}, \bibinfo {author} {\bibfnamefont {M.~A.}\ \bibnamefont
  {Burgos}}, \bibinfo {author} {\bibfnamefont {C.}~\bibnamefont {del Pino}},
  \bibinfo {author} {\bibfnamefont {M.~A.}\ \bibnamefont {Sevilla-Garcia}}, \
  and\ \bibinfo {author} {\bibfnamefont {F.}~\bibnamefont {Esteban-Ortega}},\
  }\bibfield  {title} {\enquote {\bibinfo {title} {{Robust nondimensional
  estimators to assess the nasal airflow in health and disease}},}\ }\href
  {\doibase 10.1002/cnm.2906} {\bibfield  {journal} {\bibinfo  {journal}
  {International Journal for Numerical Methods in Biomedical Engineering}\ }
  (\bibinfo {year} {2018}),\ 10.1002/cnm.2906}\BibitemShut {NoStop}%
\bibitem [{\citenamefont {Garcia}\ \emph {et~al.}(2016)\citenamefont {Garcia},
  \citenamefont {Hariri}, \citenamefont {Patel},\ and\ \citenamefont
  {Rhee}}]{Garcia2016}%
  \BibitemOpen
  \bibfield  {author} {\bibinfo {author} {\bibfnamefont {G.~J.}\ \bibnamefont
  {Garcia}}, \bibinfo {author} {\bibfnamefont {B.~M.}\ \bibnamefont {Hariri}},
  \bibinfo {author} {\bibfnamefont {R.~G.}\ \bibnamefont {Patel}}, \ and\
  \bibinfo {author} {\bibfnamefont {J.~S.}\ \bibnamefont {Rhee}},\ }\bibfield
  {title} {\enquote {\bibinfo {title} {{The relationship between nasal
  resistance to airflow and the airspace minimal cross-sectional area}},}\
  }\href {\doibase 10.1016/j.jbiomech.2016.03.051} {\bibfield  {journal}
  {\bibinfo  {journal} {Journal of Biomechanics}\ } (\bibinfo {year} {2016}),\
  10.1016/j.jbiomech.2016.03.051}\BibitemShut {NoStop}%
\bibitem [{\citenamefont {Yu}\ \emph {et~al.}(2008)\citenamefont {Yu},
  \citenamefont {Liu}, \citenamefont {Sun},\ and\ \citenamefont {S}}]{Yu2008}%
  \BibitemOpen
  \bibfield  {author} {\bibinfo {author} {\bibfnamefont {S.}~\bibnamefont
  {Yu}}, \bibinfo {author} {\bibfnamefont {L.}~\bibnamefont {Liu}}, \bibinfo
  {author} {\bibfnamefont {X.}~\bibnamefont {Sun}}, \ and\ \bibinfo {author}
  {\bibfnamefont {L.}~\bibnamefont {S}},\ }\bibfield  {title} {\enquote
  {\bibinfo {title} {Influence of nasal structure on the distribution of
  airflow in nasal cavity},}\ }\href {\doibase PMID18575016.} {\bibfield
  {journal} {\bibinfo  {journal} {Rhinology}\ }\textbf {\bibinfo {volume}
  {46}},\ \bibinfo {pages} {137--143} (\bibinfo {year} {2008})}\BibitemShut
  {NoStop}%
\bibitem [{\citenamefont {Hildebrandt}\ \emph {et~al.}(2013)\citenamefont
  {Hildebrandt}, \citenamefont {Goubergrits}, \citenamefont {Heppt},
  \citenamefont {Bessler},\ and\ \citenamefont {Zachow}}]{Hildebrandt2013}%
  \BibitemOpen
  \bibfield  {author} {\bibinfo {author} {\bibfnamefont {T.}~\bibnamefont
  {Hildebrandt}}, \bibinfo {author} {\bibfnamefont {L.}~\bibnamefont
  {Goubergrits}}, \bibinfo {author} {\bibfnamefont {W.~J.}\ \bibnamefont
  {Heppt}}, \bibinfo {author} {\bibfnamefont {S.}~\bibnamefont {Bessler}}, \
  and\ \bibinfo {author} {\bibfnamefont {S.}~\bibnamefont {Zachow}},\
  }\bibfield  {title} {\enquote {\bibinfo {title} {{Evaluation of the
  intranasal flow field through computational fluid dynamics}},}\ }\href
  {\doibase 10.1055/s-0033-1341591} {\bibfield  {journal} {\bibinfo  {journal}
  {Facial Plastic Surgery}\ } (\bibinfo {year} {2013}),\
  10.1055/s-0033-1341591}\BibitemShut {NoStop}%
\end{thebibliography}%

\normalsize
\newgeometry{margin=2.0cm}
\begin{figure}
	\centering
	\includegraphics[width=0.85\linewidth]{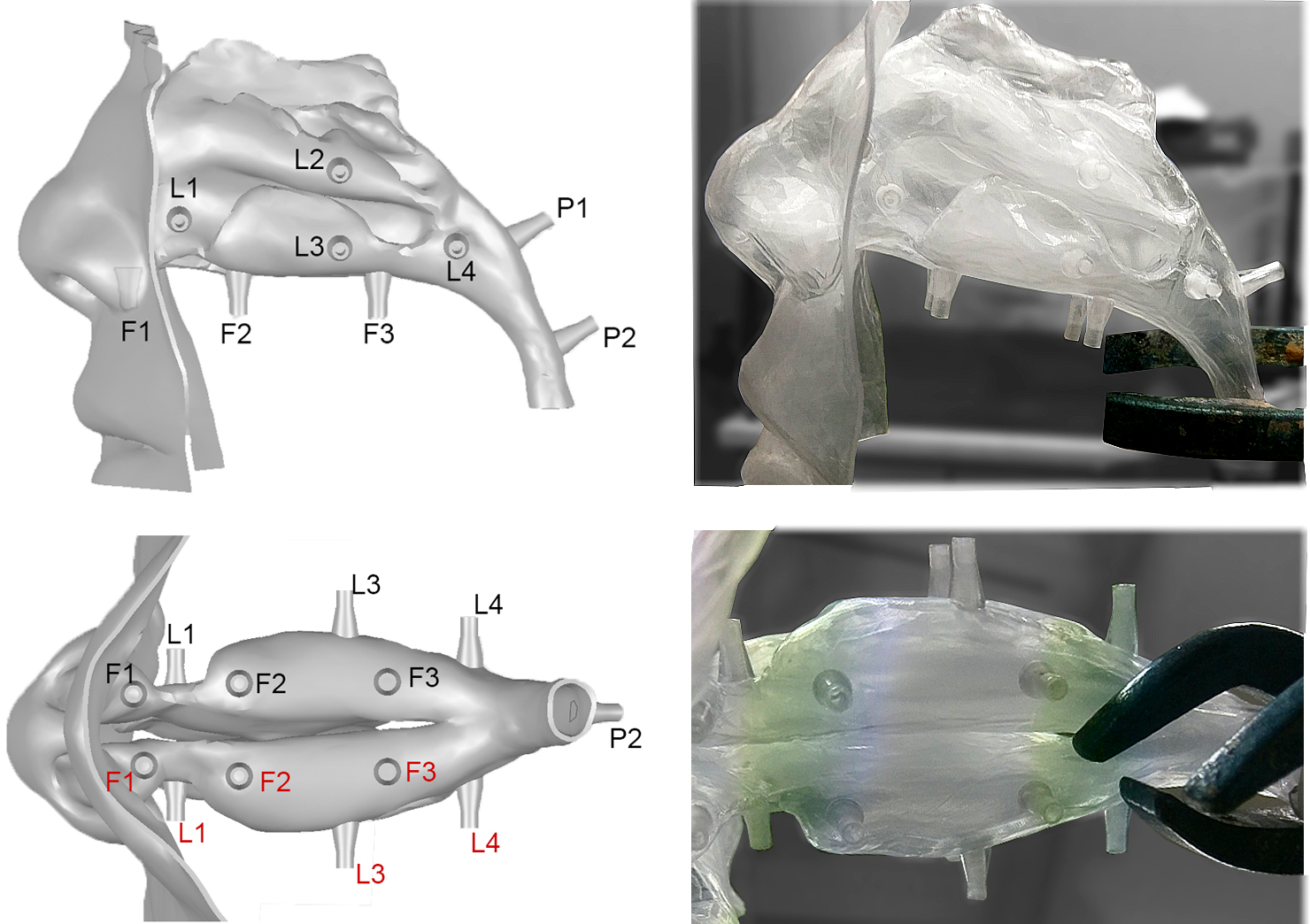}
	\caption{Labelled schematic of the computational model showing the port locations where the pressure transducer was placed, and its corresponding photo image. The port locations were placed symmetrically between the left and right cavity, and labelled with $L$ for locations on the lateral walls; and $F$ for locations on the nasal floor. Location $F1$ sits along the floor of the vestibule and is hidden behind the face in the image. The two posterior ports after the left and right cavities had merged were labelled as $P1$ and $P2$.}
	\label{fig:portlocations}
\end{figure}
\clearpage

\begin{figure}
	\centering
	\includegraphics[width=0.85\linewidth]{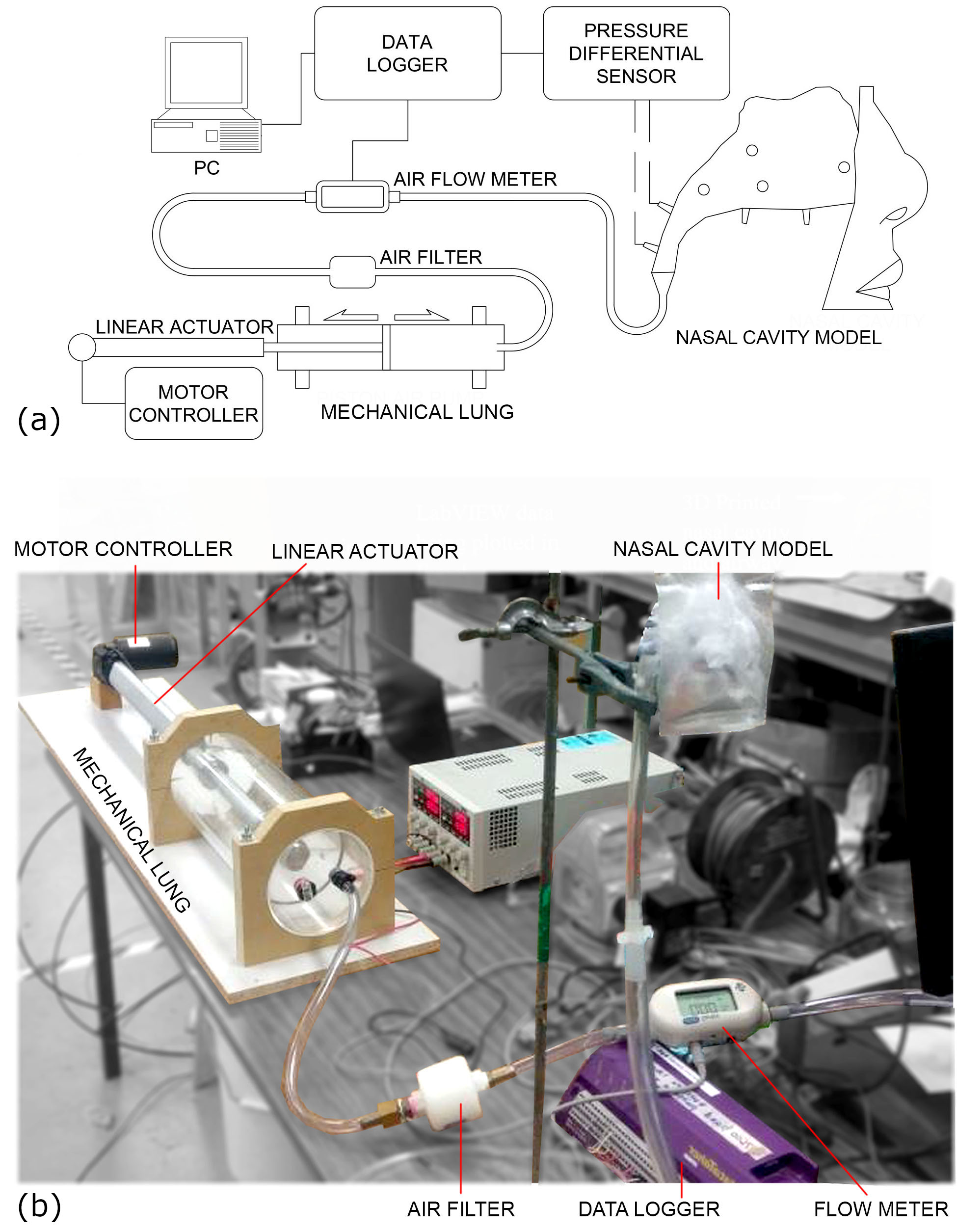}
	\caption{Schematic and photo of the experimental test rig setup used for pressure differential measurements.}
	\label{fig:pressuredifferential}
\end{figure}
\clearpage

\begin{figure}
	\centering
	\begin{subfigure}[b]{0.75\textwidth}
		\includegraphics[width=\textwidth]{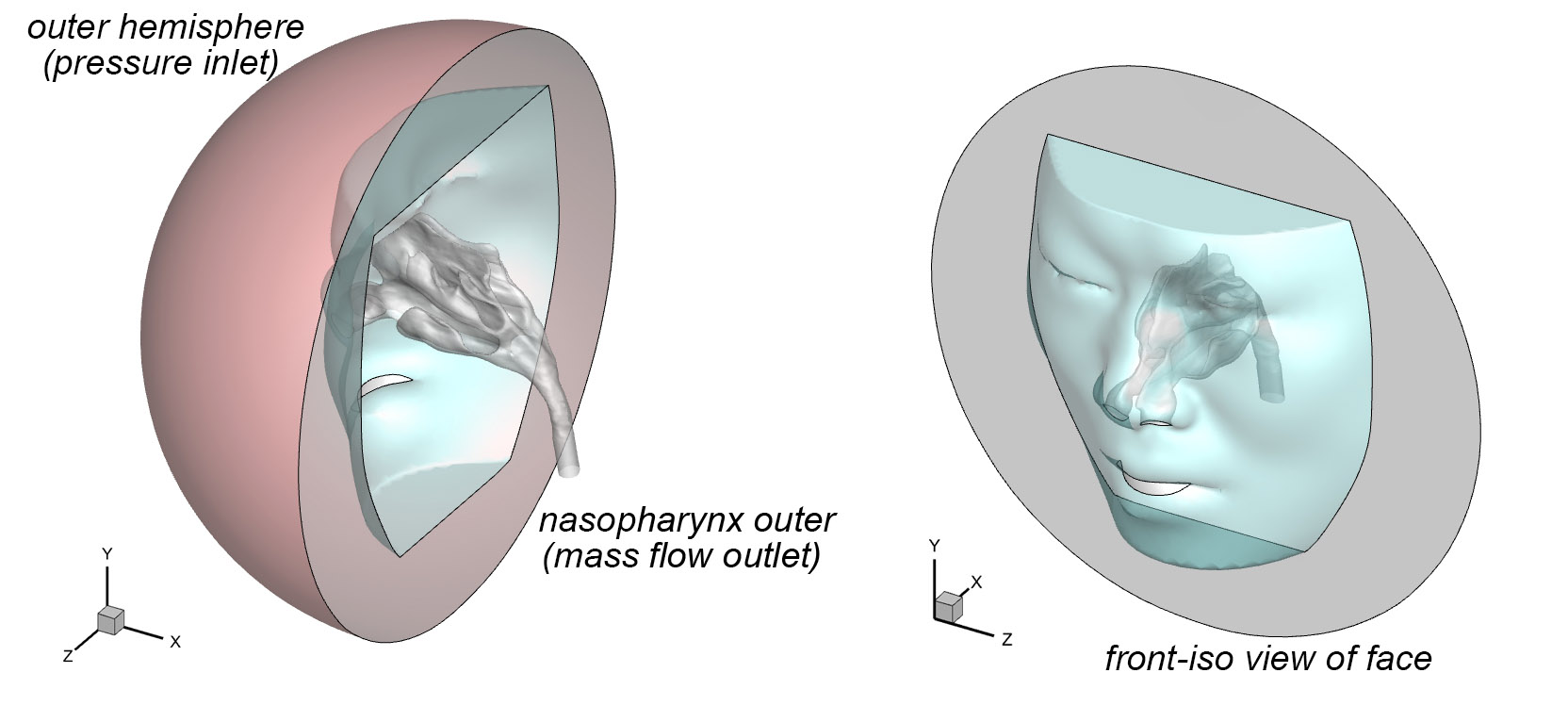}
		\caption{Computational domain}
		\vspace*{4mm}
	\end{subfigure}
	~
	\begin{subfigure}[b]{0.75\textwidth}
		\includegraphics[width=\textwidth]{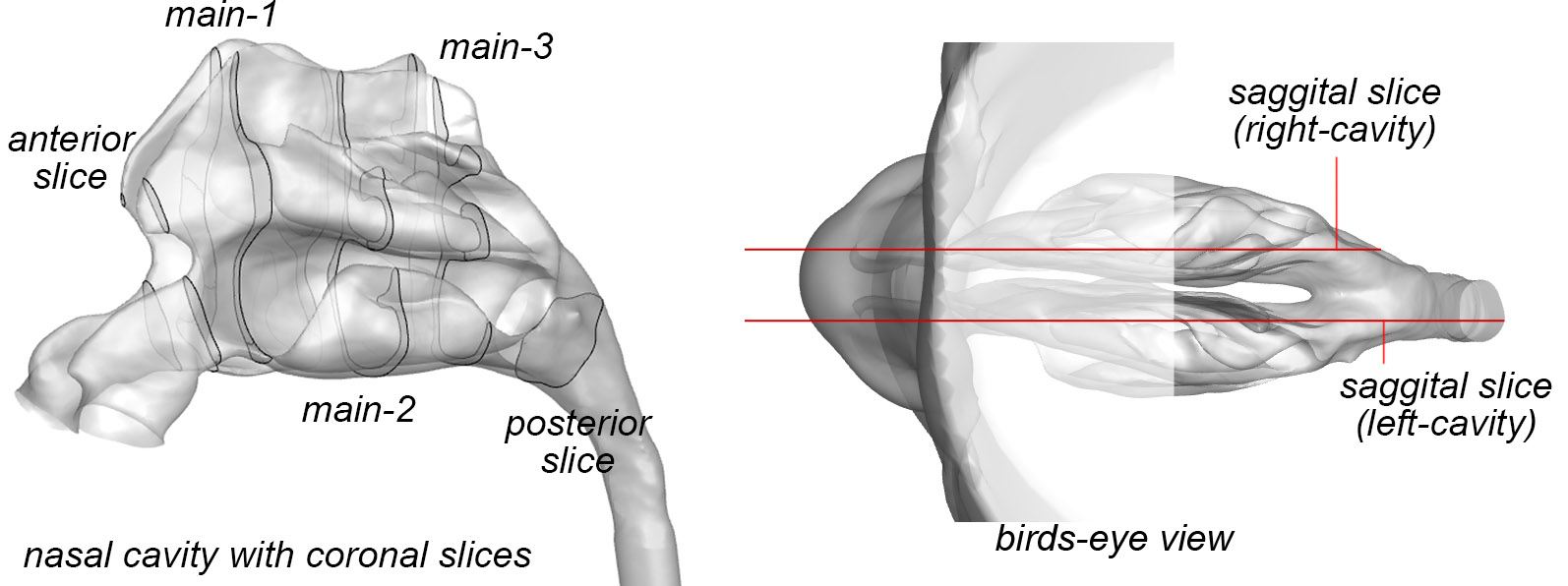}
		\caption{Coronal and sagittal slices}
		\vspace*{4mm}
	\end{subfigure}
	~
	\begin{subfigure}[b]{0.75\textwidth}
		\includegraphics[width=\textwidth]{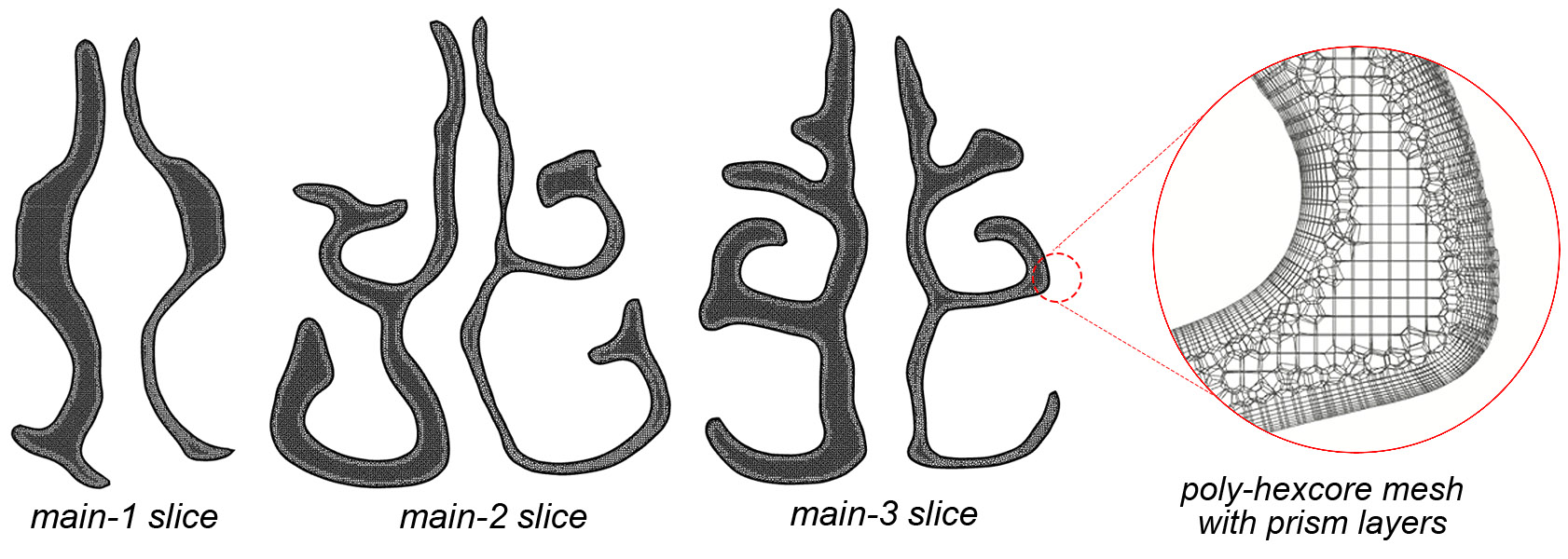}
		\caption{Internal mesh}
	\end{subfigure}
	
	\caption{Nasal cavity geometry showing the computational domain that includes (a) the outer face and an outer hemisphere. Planar slices were created for post-processing steps.  The nasal cavity shown with coronal slice locations. The slices cutting through the anterior, main passage, and posterior locations are shown. Two sagittal slices were cut through each nasal cavity (left and right) side. (c) The internal mesh of main-1, main-2, main-3 are shown, and a zoomed view of the mesh shows 8-prism layers, and internal hexahedral cell with length 0.2~mm, i.e. volume =  $8.0\times 10^{-12}$~m$^3$}
	\label{fig:geom}
\end{figure}

\clearpage
\begin{figure}
	\centering
	\begin{subfigure}[b]{1\textwidth}
		\includegraphics[width=1\textwidth]{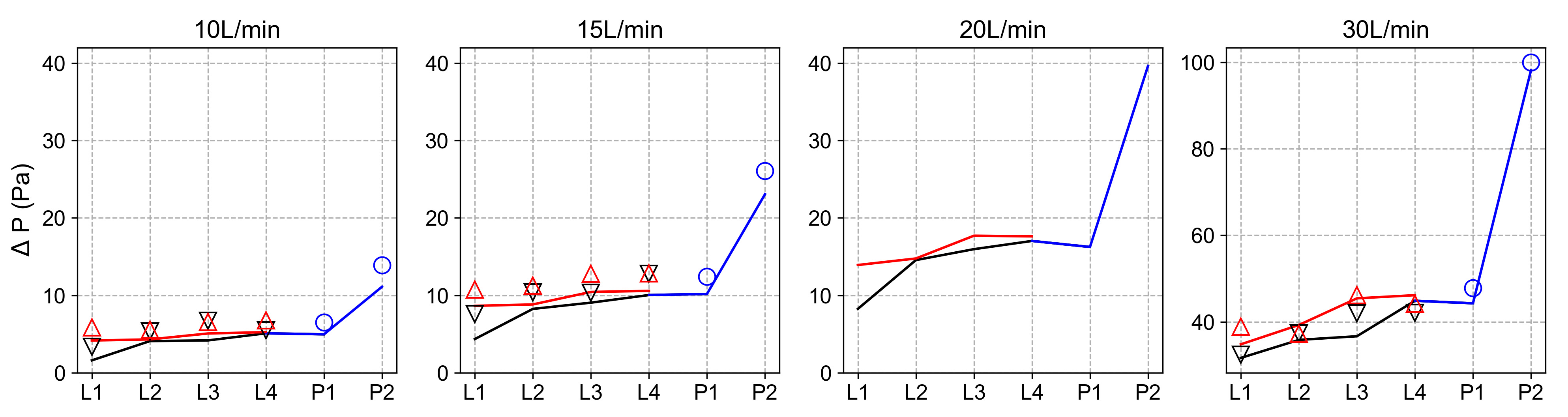}
		\caption{Lateral nasal port locations}
		\vspace*{4mm}
	\end{subfigure}
	
	\begin{subfigure}[b]{1\textwidth}
		\includegraphics[width=\textwidth]{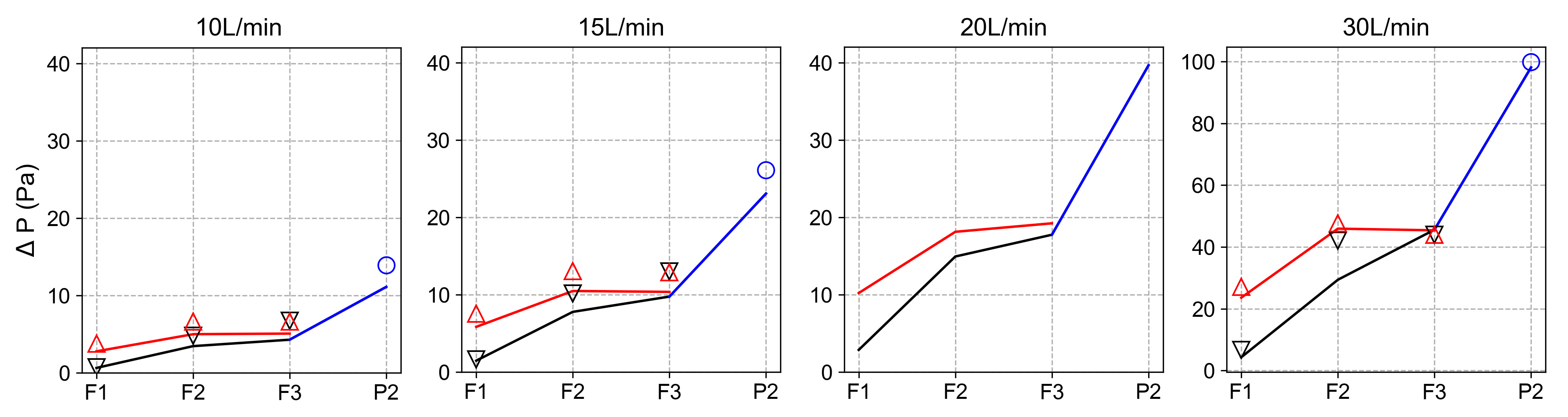}
		\caption{Nasal floor port locations}
		\vspace*{4mm}
	\end{subfigure}
	
	\begin{subfigure}[b]{0.65\textwidth}
		\centering
		\includegraphics[width=0.65\textwidth]{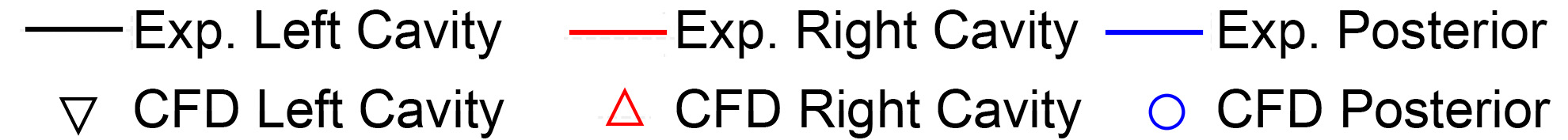}
		\vspace*{4mm}
	\end{subfigure}
	
	\caption{Pressure differential between ambient surrounding air and port locations (a) at the lateral walls labelled L1 to L4; and (b) along the nasal floor labelled F1 to F3 (see Fig.~\ref{fig:portlocations} for location definitions). The lines and markers are coloured by airway side, where black = left cavity; red = right cavity; blue is the posterior region where there is a single passageway only. CFD simulations were  performed for flow rates of 10, 15, 20 and 30~L/min.}
	\label{fig:press-distn}
\end{figure}
\clearpage

\begin{figure}
	\centering
	\includegraphics[width=0.7\linewidth]{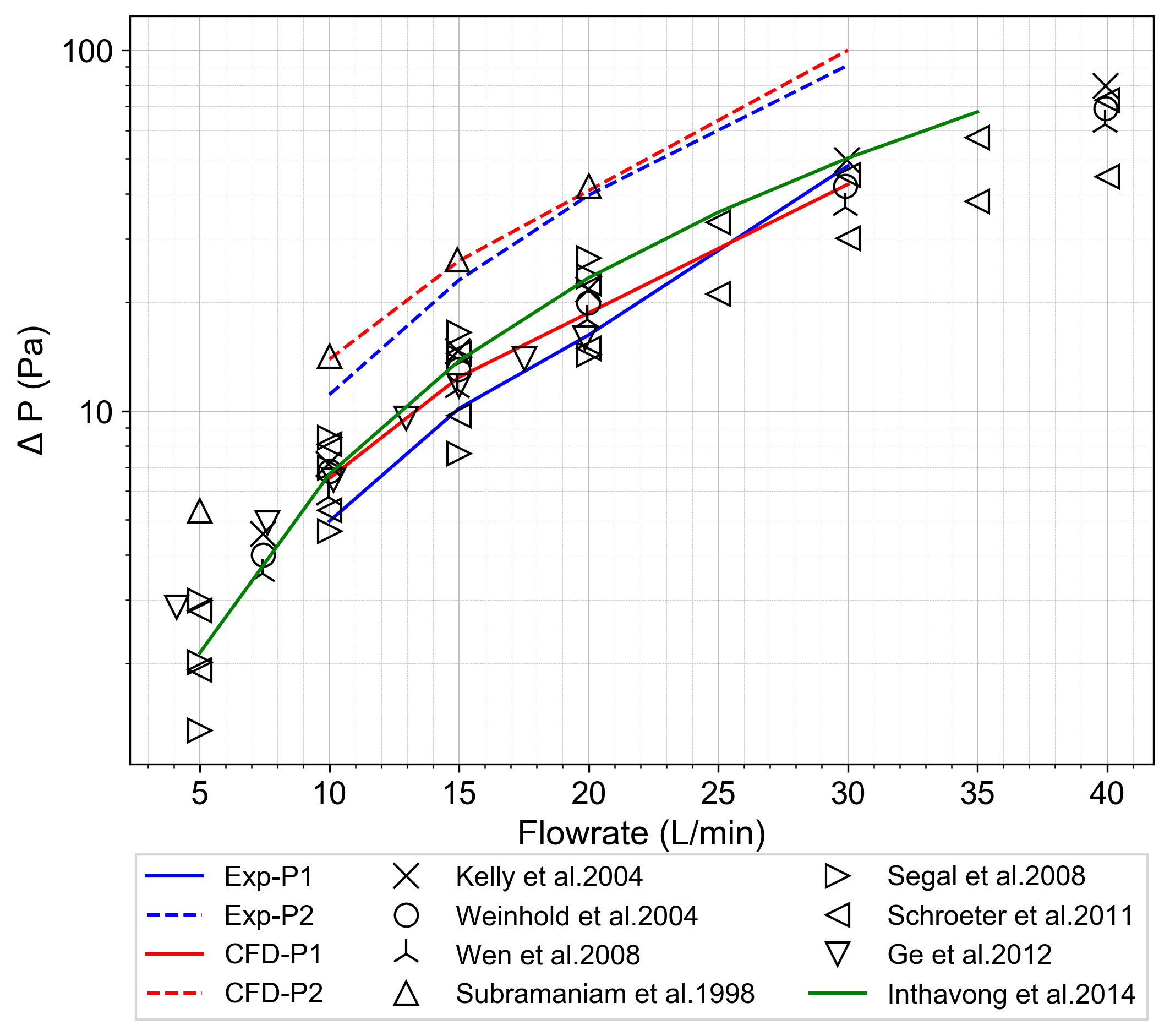}
	\caption{Overall pressure drop from the ambient air and posterior locations P1, and P2 (see Fig.~\ref{fig:portlocations} for location definitions) compared with literature data \citep{Weinhold2004, Kelly2004, Wen2008, Subramaniam1998, Segal2008, Ge2012}. Line data of Inthavong et al.\cite{Inthavong2014} uses the equation $\Delta P = (0.2033\sqrt{\rho}A^{-1}Q + 0.338\sqrt{\rho}A^{-1}Q_0)^2$ where $\rho = 1.225$kg/m$^3$, $A$ is per unit area in m$^2$, $Q$ is the flowrate, and $Q_0$ is a unit flow rate}
	\label{fig:pressDrop}
\end{figure}
\clearpage

\begin{figure}
	\centering
	\begin{subfigure}[b]{0.6\textwidth}
		\includegraphics[width=\textwidth]{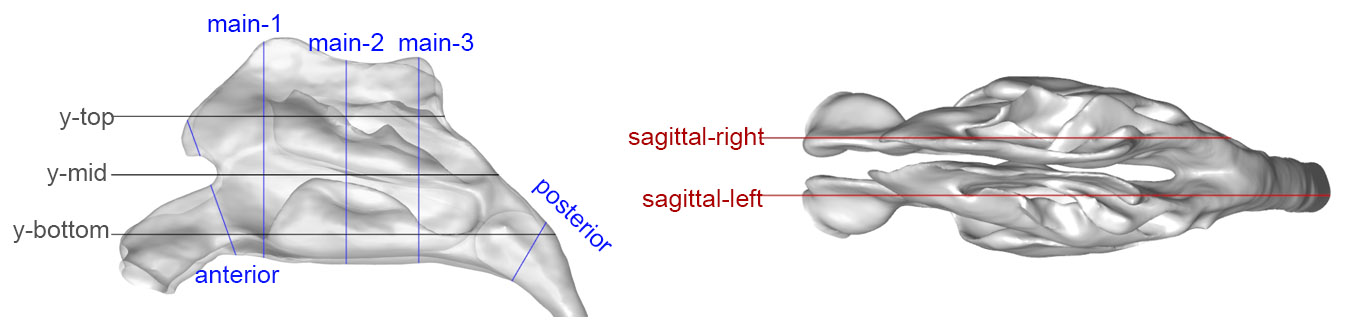}
		\caption{All slice locations, e.g., coronal, sagittal, and transverse}
		\vspace*{2mm}
	\end{subfigure}
	~
	
	\begin{subfigure}[b]{0.65\textwidth}
		\includegraphics[width=\textwidth]{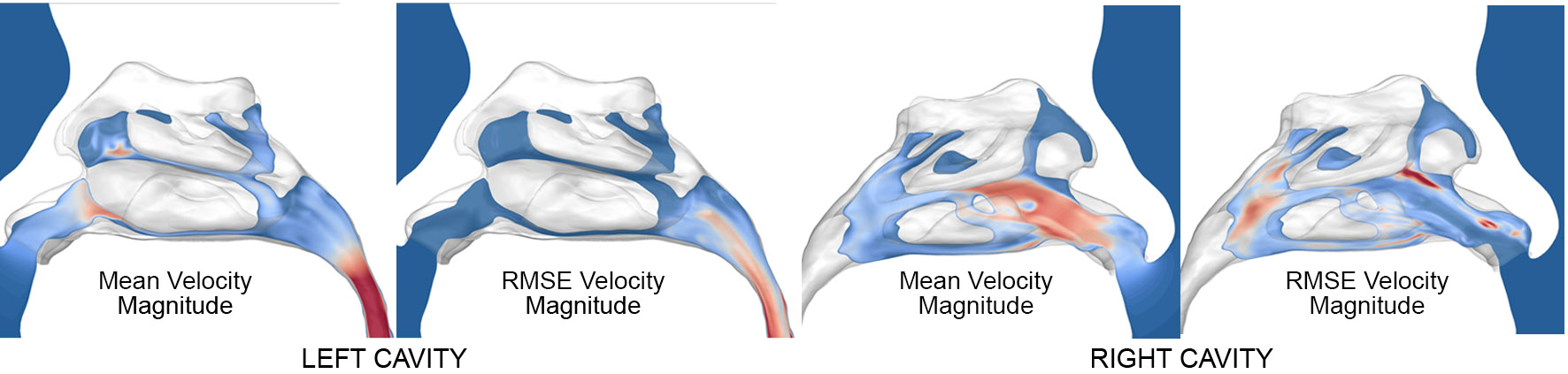}
		\caption{Velocity contours in the sagittal planes}
		\vspace*{2mm}
	\end{subfigure}
	~
	\begin{subfigure}[b]{0.725\textwidth}
		\includegraphics[width=\textwidth]{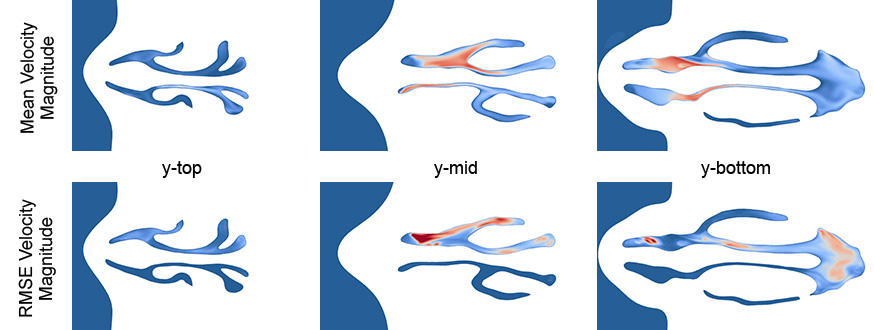}
		\caption{Velocity contours in the transverse planes}
		\vspace*{2mm}
	\end{subfigure}
	~
	\begin{subfigure}[b]{0.75\textwidth}
		\includegraphics[width=\textwidth]{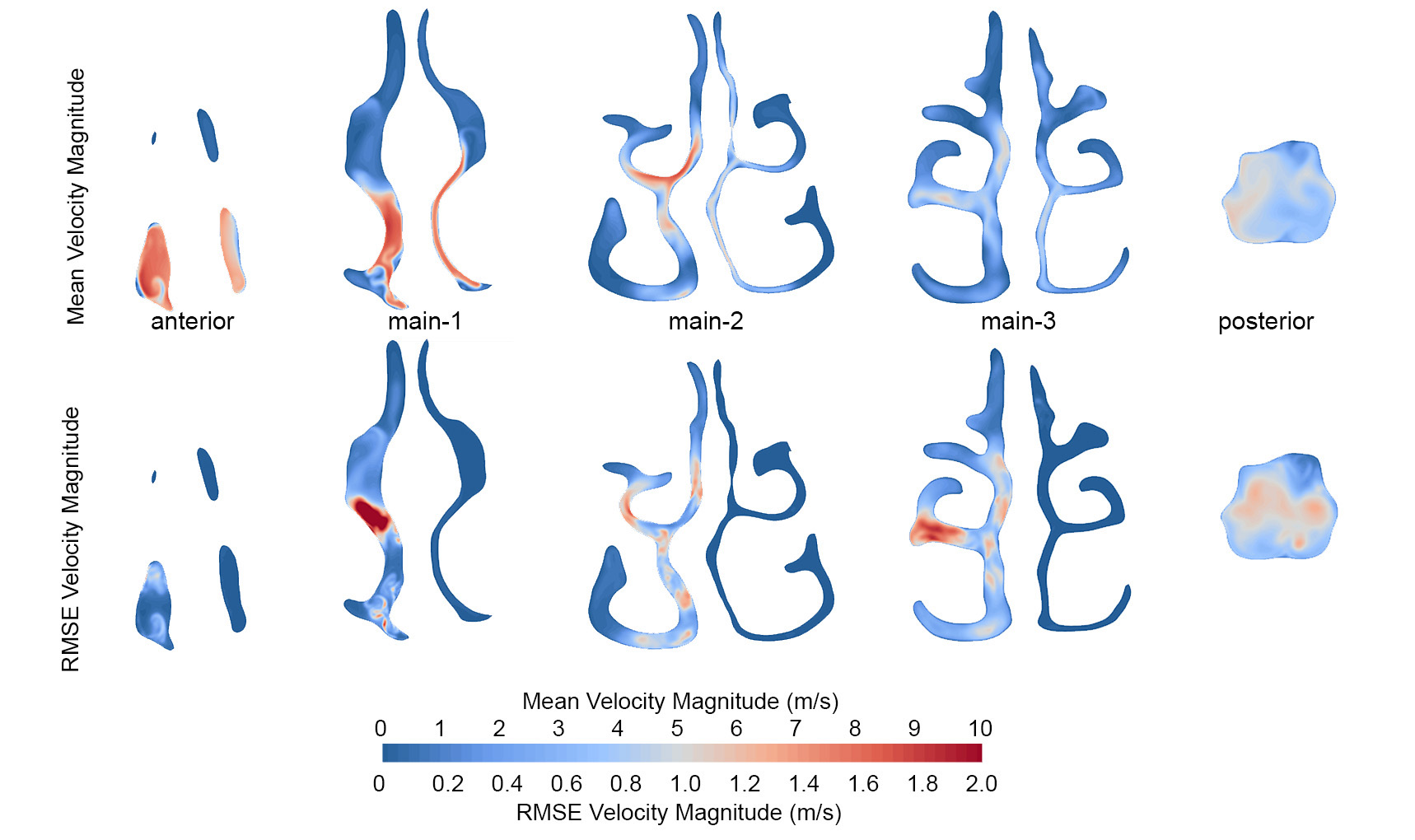}
		\caption{Velocity contours in the coronal planes}
		\vspace*{2mm}
	\end{subfigure}
	
	\caption{(a) Cross-sectional plane locations and (b) Mean and RMSE velocity contours shown at the sagittal (c) transverse, and (d) coronal planes.}
	\label{fig:meanVals}
\end{figure}
\clearpage

\begin{figure}
	\centering
	\begin{subfigure}[b]{1\textwidth}
		\includegraphics[width=0.95\linewidth]{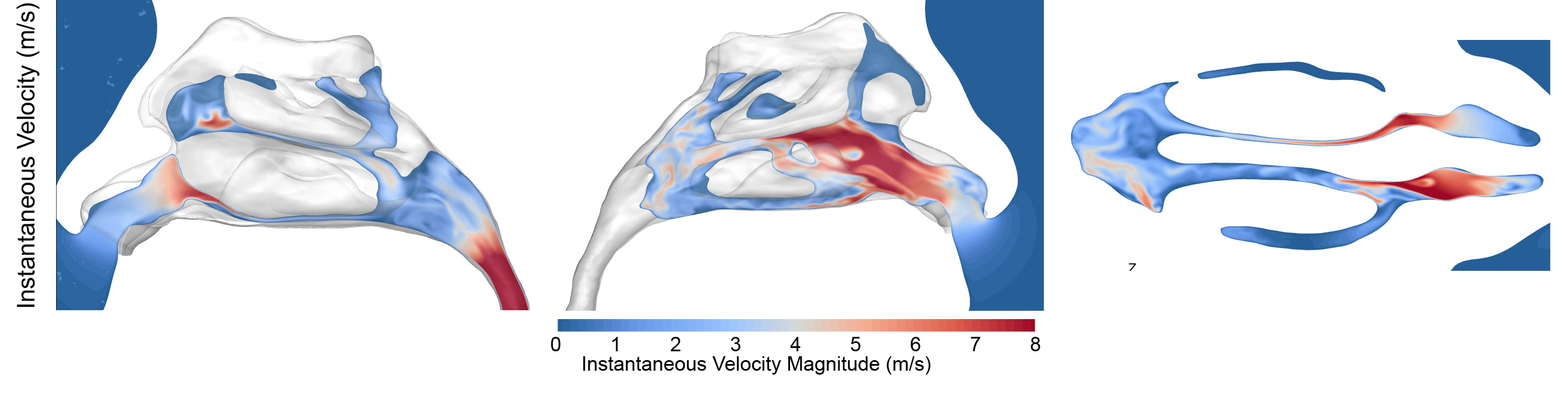}
		\caption{Instantaneous velocity contours}
		\vspace*{2mm}
	\end{subfigure}
	~
	\begin{subfigure}[b]{1\textwidth}
		\includegraphics[width=0.95\linewidth]{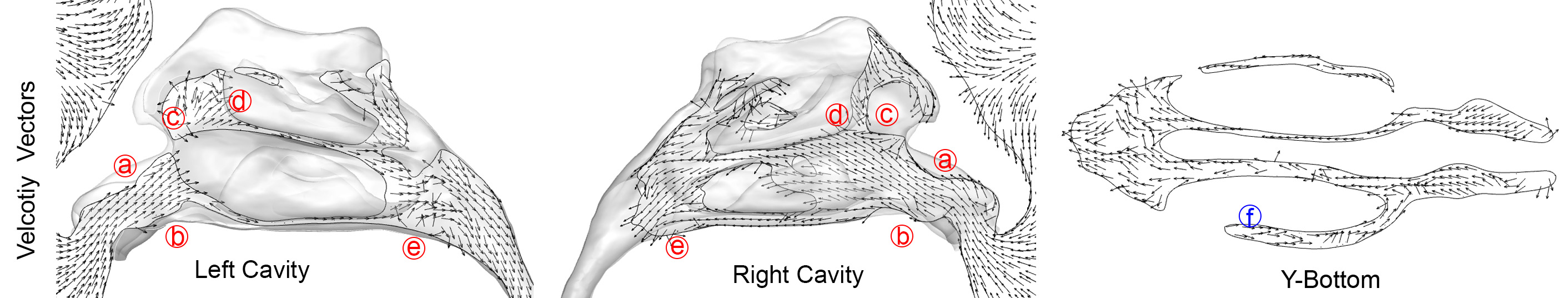}
		\caption{Instantaneous velocity vectors}
		\vspace*{2mm}
	\end{subfigure}
	
	\caption{(a) Instantaneous velocity contours taken on the sagittal and transverse planes, and (b) velocity vectors where labels circled in red highlight key flow characteristics.}
	\label{fig:flowfea}
\end{figure}
\clearpage

\begin{figure}
	\centering
	\begin{subfigure}[b]{1\textwidth}
		\includegraphics[width=0.95\linewidth]{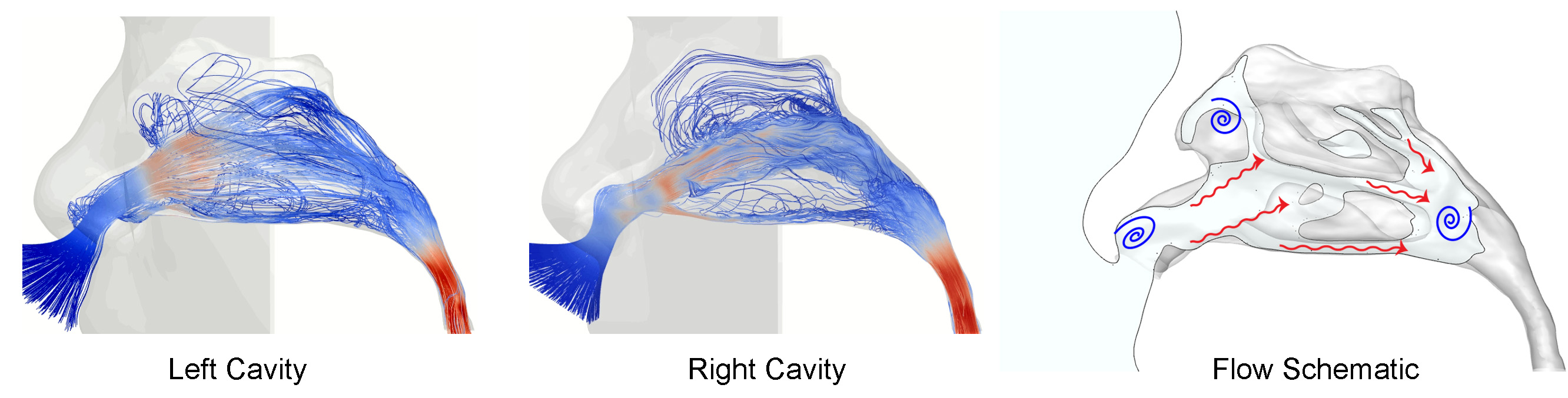}
		\caption{Streamlines and an overall flow schematic of the CFD results}
		\vspace*{2mm}
	\end{subfigure}
	~
	\begin{subfigure}[b]{1\textwidth}
		\includegraphics[width=0.95\linewidth]{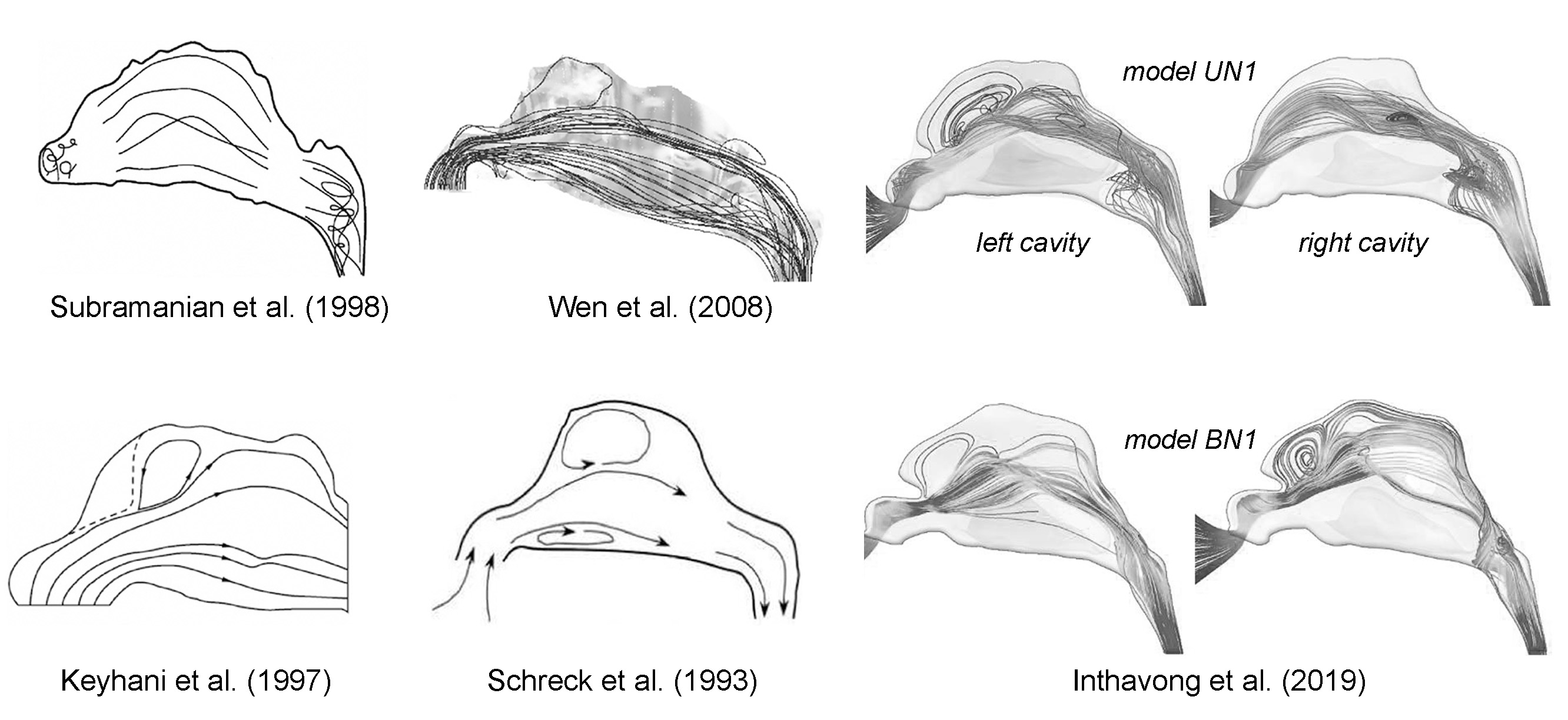}
		\caption{Comparative streamlines reported in the literature}
		\vspace*{2mm}
	\end{subfigure}
	
	\caption{(a) Pathlines in the left and right cavity showing the flow behaviour and identifying recirculating flow regions, and a schematic summarising the main flow features. (b) Reported streamline schematics or CFD simulation results found in the literature for comparison.}
	\label{fig:stream}
\end{figure}
\clearpage

\begin{figure}
	\centering
	\includegraphics[width=0.95\linewidth]{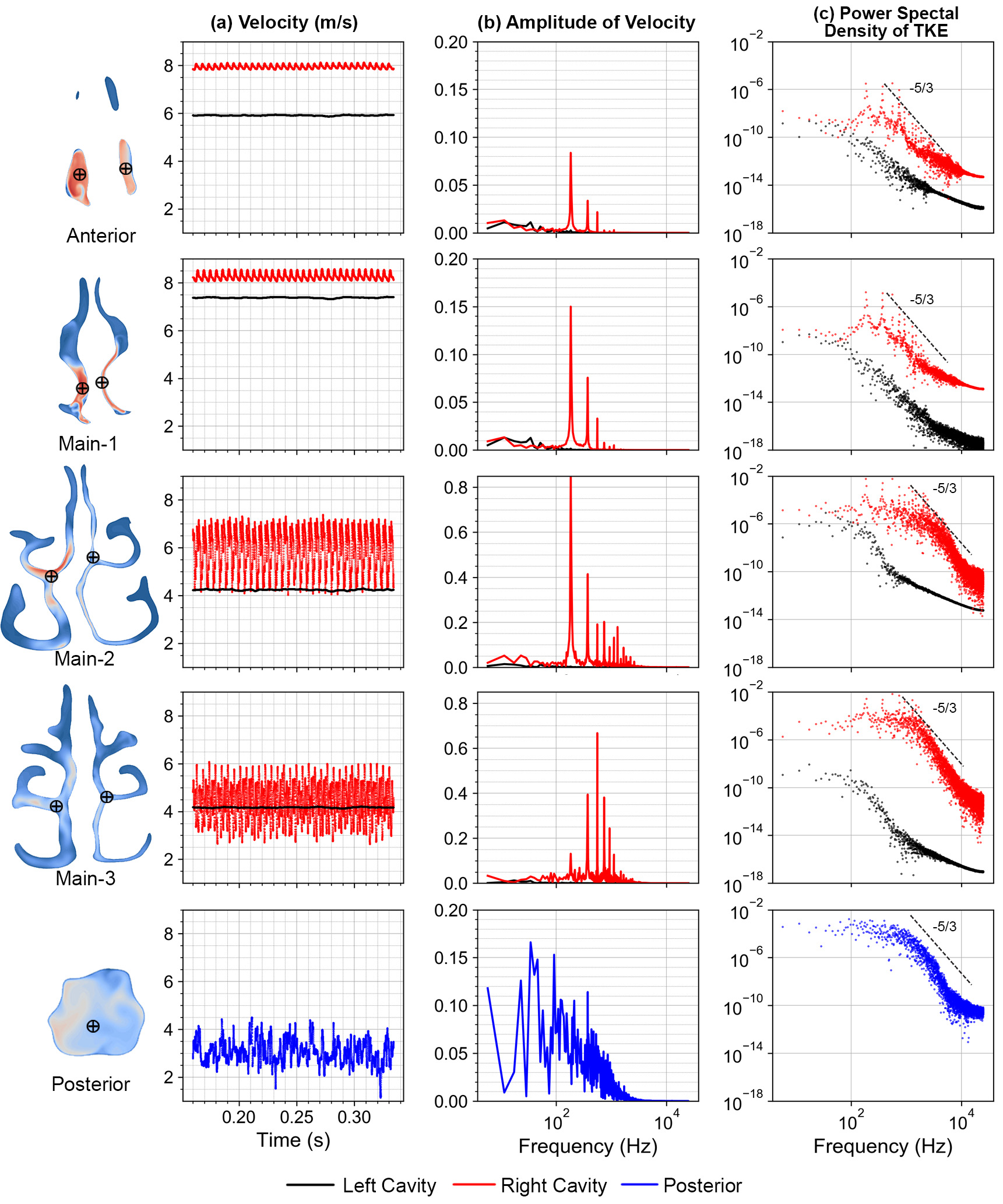}
	\caption{Transient monitor data for (a) velocity magnitude, (b) spectral analysis of the velocity, and (c) spectral analysis of turbulence kinetic energy. The monitoring points were located in the coronal planes and monitored over the time period $t = 1.6 $ to $2.0$~s. The turbulence kinetic energy, $k = k_r + k_u$, was obtained from the resolved component, $k_r = 0.5(\overline{u^{'2}}+\overline{v^{'2}}+\overline{w^{'2}})$ ) and unresolved component, $k_u = (C_w \Delta S)^2 / 0.3$, where $\Delta$ is the grid length scale and $C_w = 0.325$ in the WALE model.}
	\label{fig:fluct}
\end{figure}

\begin{figure}
	\centering
	\includegraphics[width=0.95\linewidth]{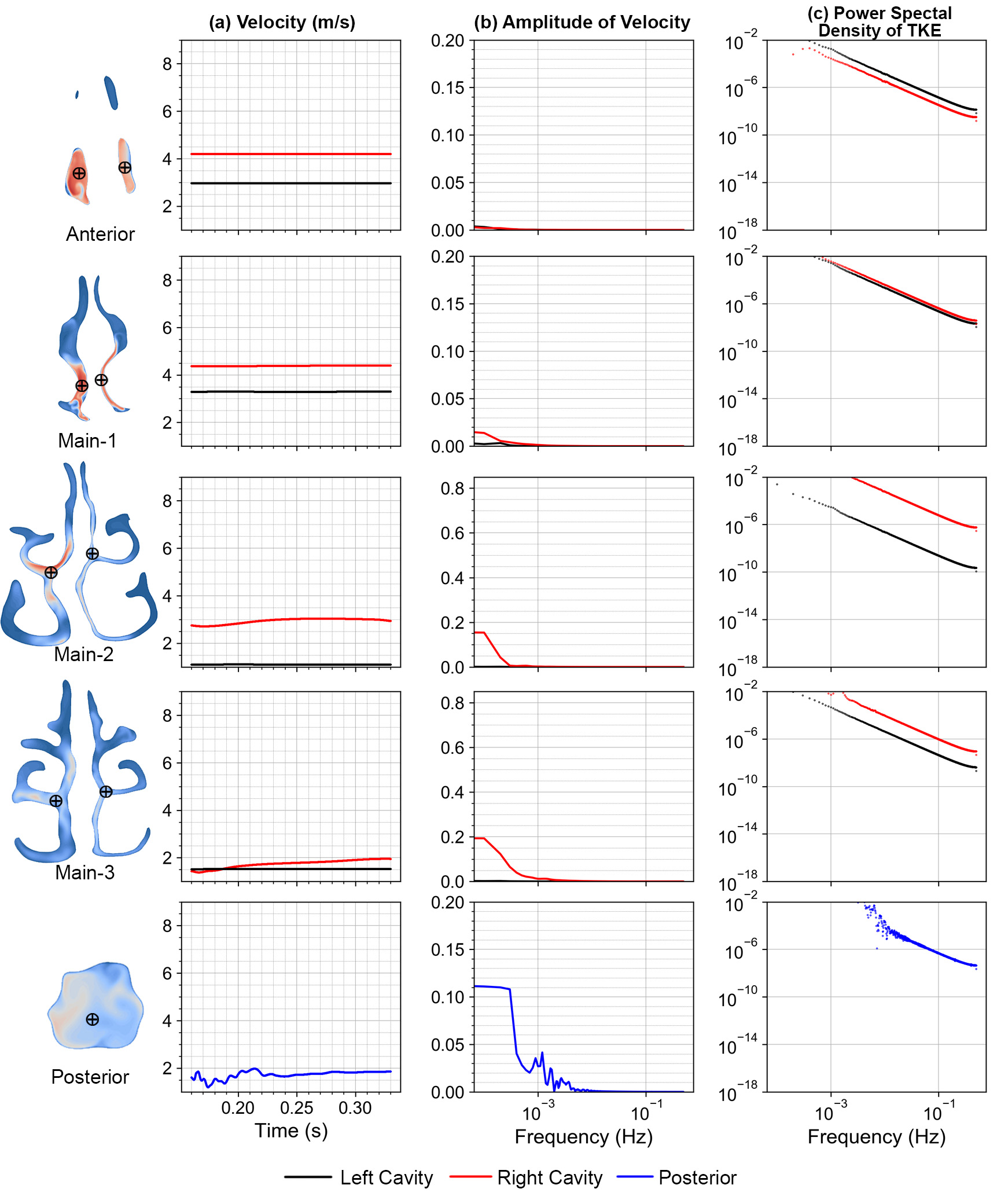}
	\caption{Transient monitor data at the same locations and setup as in Fig.~\ref{fig:fluct}, except for the flow rate which is now 15~L/min. (a) velocity magnitude, (b) spectral analysis of the velocity, and (c) spectral analysis of turbulence kinetic energy }
	\label{fig:fluct15}
\end{figure}

\clearpage
\section*{Supplementary Material}
\subsection*{S1: Velocity Fluctuation Animation}
\begin{enumerate}
	\item Animation 1: Coronal Planes  (see attached file)
	\item Animation 2: Sagittal Planes (see attached file)
\end{enumerate}

\clearpage
\subsection*{S2: Pressure Distribution Data}
\begin{table}[!h]
	\centering
	\small 
	\begin{tabular}{|l|l|l|l|l|l|l|l|}
		\hline
		& \multicolumn{4} {c|} {Experimental, Flow Rates}  & \multicolumn{3} {c} {CFD, Flow Rates}  \\ \hline
		Location & 10 L/min & 15 L/min & 20 L/min & 30 L/min & 10 L/min & 15 L/min & 30 L/min \\ \hline
		L1 Left & 1.59 & 4.35 & 8.24 & 31.64 & 3.33 & 7.59 & 32.43 \\ \hline
		L2 Left& 4.08 & 8.24 & 14.55 & 35.81 & 5.34 & 10.37 & 37.33 \\ \hline
		L3 Left & 4.17 & 9.05 & 15.96 & 36.65 & 6.70 & 10.36 & 41.99 \\ \hline
		L4 Left & 5.09 & 10.04 & 17.03 & 44.85 & 5.51 & 12.76 & 42.03 \\ \hline
		P1 & 4.96 & 10.17 & 16.25 & 44.27 & 6.52 & 12.42 & 47.78 \\ \hline
		P2 & 11.11 & 23.07 & 39.64 & 98.15 & 13.93 & 26.09 & 99.95 \\ \hline
		L1 Right & 4.16 & 8.64 & 13.92 & 34.79 & 5.84 & 10.73 & 38.87 \\ \hline
		L2 Right & 4.30 & 8.82 & 14.76 & 39.26 & 5.55 & 11.24 & 37.20 \\ \hline
		L3 Right & 5.07 & 10.44 & 17.70 & 45.42 & 6.56 & 12.79 & 46.17 \\ \hline
		L4 Right & 5.24 & 10.57 & 17.62 & 46.15 & 6.77 & 12.85 & 44.20 \\ \hline
		F1 Left & 0.63 & 1.47 & 2.86 & 4.28 & 0.70 & 1.61 & 6.70 \\ \hline
		F2 Left & 3.45 & 7.78 & 14.93 & 29.37 & 4.81 & 10.14 & 41.99 \\ \hline
		F3 Left & 4.28 & 9.75 & 17.75 & 45.58 & 6.75 & 13.01 & 43.93 \\ \hline
		F1 Right & 2.77 & 5.84 & 10.21 & 23.58 & 3.79 & 7.52 & 27.14 \\ \hline
		F2 Right & 4.99 & 10.47 & 18.12 & 45.93 & 6.61 & 13.05 & 47.62 \\ \hline
		F3 Right & 5.07 & 10.36 & 19.22 & 45.38 & 6.65 & 12.88 & 44.00 \\ \hline
		\multicolumn{8} {|c|} {Pressure Drop Contributions}  \\ \hline
		\multicolumn{8} {|c|} {\textit{Nasal Valve (F2-F1) } Pa} \\ \hline
		Left & 2.8 Pa & 6.3 Pa & 12.1 Pa & 25.1 Pa & 4.1 Pa & 8.5 Pa & 35.3 Pa \\ \hline
		Right & 2.2 Pa & 4.2 Pa & 6.1 Pa & 20.8 Pa & 2.5 Pa & 4.5 Pa & 12.3 Pa \\ \hline
		
		\multicolumn{8} {|c|} {\textit{Nasal Valve (F2-F1) to P1 Contribution}\%}  \\ \hline
		Left & 56.8\% & 62.0\% & 74.3\% & 56.7\% & 63.1\% & 68.6\% & 73.9\% \\ \hline
		Right & 43.9\% & 40.9\% & 37.3\% & 47.1\% & 38.3\% & 36.4\% & 25.8\% \\ \hline
		
	\end{tabular}
	\caption{Pressure distribution values extracted from experimental measurements and CFD simulation results}
\end{table}

\subsection*{S3: CAD geometry of the nasal cavity for 3D printing}
A print-ready stl file of the 3D nasal cavity model with pressure ports depicted in Fig \ref{fig:geom}.

\subsection*{S4: CAD geometry of the nasal cavity used for CFD analysis}
An STL file of the 3D nasal cavity model ready for CFD meshing.

\end{document}